\documentclass[5p]{elsarticle}

\usepackage{hyperref,color}
\usepackage[switch]{lineno}
\modulolinenumbers[5]

\journal{Astroparticle Physics}
\usepackage{amsmath}
\usepackage{multirow} 
\usepackage{morefloats}  
\usepackage{textcomp} 
\biboptions{sort&compress}
\usepackage[utf8]{inputenc}
\usepackage{dcolumn}
\newcolumntype{d}[1]{D{$\pm$}{$\pm$}{#1}}    
\usepackage{tablefootnote}

\usepackage{upgreek}
\usepackage{footnote}
\usepackage{amssymb} 
\usepackage{siunitx}

\begin{document}
\begin{frontmatter}	
\title{A new ultra low-level HPGe activity counting setup in the Felsenkeller shallow-underground laboratory}	
		
\author[a]{S. Turkat}
\author[b]{D. Bemmerer}
\ead{d.bemmerer@hzdr.de}
\author[b]{A. Boeltzig}
\author[a]{A. R. Domula}
\author[a,b]{J. Koch}
\author[a,b]{T. Lossin}
\author[a,b]{M. Osswald}
\author[b]{K. Schmidt}
\author[a]{K. Zuber}
\ead{kai.zuber@tu-dresden.de}

\address[a]{Technische Universität Dresden (TU Dresden), 01069 Dresden, Germany}		
\address[b]{Helmholtz-Zentrum Dresden-Rossendorf (HZDR), Bautzner Landstr. 400, 01328 Dresden, Germany}
		
\begin{abstract}
A new ultra low-level counting setup has been installed in the shallow-underground laboratory Felsenkeller in Dresden, Germany. It includes a high-purity germanium detector (HPGe) of 163\,\% relative efficiency within passive and active shields. The passive shield consists of 45m rock overburden (140 meters water equivalent), 40\,cm of low-activity concrete, and a lead and copper castle enclosed by an anti-radon box. The passive shielding alone is found to reduce the background rate to rates comparable to other shallow-underground laboratories. An additional active veto is given by five large plastic scintillation panels surrounding the setup. It further reduces the background rate by more than one order of magnitude down to 116(1)\,kg$^{-1}$d$^{-1}$ in an energy interval of [40\,keV;2700\,keV]. This low background rate is unprecedented for shallow-underground laboratories and close to deep underground laboratories.
\end{abstract}
		
\date{}	
\end{frontmatter}

\section{Introduction}
\label{sec:Introduction}

A variety of scientific fields call for detection systems which are able to measure radioactive samples with ultra-low activities in the order of $\upmu$Bq to mBq \cite{Povinec07-Book}. A case in point is the search for rare processes such as dark matter interactions \cite{SuperCDMS2017, Edelweiss2017} or rare germanium decays \cite{Zuber2021}. These include neutrino-accompanied (2$\nu\beta\beta$) \cite{Brudanin2017} and neutrinoless (0$\nu\beta\beta$) \cite{Alvis2019,2020PhRvL.125y2502A} double beta decays, among others. 

The need for ultra-low background radioactivity measurements is served by a number of low-background laboratories which are using high-purity germanium (HPGe) detectors in shallow or deep underground settings \cite{Laubenstein2004}. All of these laboratories use massive lead and copper shields that are sometimes complemented by additional materials in order to suppress $\gamma$-ray background from the immediate surroundings of detector and laboratory. The effects of radioactive radon gas are usually mitigated by placing detector and sample in an airtight enclosure that is continually flushed with a radon-free gas, preferably N$_2$.

Two additional background sources are more difficult to treat, namely cosmic-ray muons and neutrons. The former usually play no role in deep-underground laboratories \cite[e.g.]{Arpesella96-Apradiso}, where the muon flux is reduced by six or more orders of magnitude compared to sea level, rendering its effects negligible for low-background activity measurements. In shallow-underground laboratories, the muon flux is usually suppressed only by one or two orders of magnitude \cite[e.g.]{Ludwig2019} and may be mitigated using active veto detectors \cite[e.g.]{Szuecs2019}.

Neutrons may be produced in two ways. First, neutron generation by muon spallation on structural or shielding materials. This process usually dominates in shallow-underground laboratories \cite{Grieger2020}. The second neutron production process are ($\alpha, n$) reactions in the rock surrounding the laboratory, with the $\alpha$ provided by natural radioactivity from the natural decay chains. ($\alpha, n$) usually dominates the remaining, low, neutron flux in deep-underground laboratories \cite{Wulandari04-APP}.

An ultra-low background radioactivity counting laboratory must address all of these above mentioned background sources. In addition, it must limit the intrinsic radioactivity of the detector by material selection \cite{Koehler2009}. 

It is noted that these techniques may also benefit, among others, experimental nuclear astrophysics. Examples include the usage of activity measurement setups to count activation products such as $^7$Be \cite{Bemmerer2006}, $^{18}$F \cite{2014PhRvC..89a5803D}, or $^{44}$Ti \cite{2013PhRvC..88b5803S}, but also in-beam measurements in underground settings \cite{2018PrPNP..98...55B}. These studies are needed for a better understanding of solar fusion \cite{Orebi2021} and Big Bang Nucleosynthesis \cite{Anders2014, Mossa2020, Turkat2021}. 

The present work reports on a large, coaxial HPGe detector called TU1, which is placed in the Felsenkeller shallow-underground laboratory. This detector has been optimized for ultra-low background activity measurements. In previous work, the Felsenkeller laboratory has already been characterized in several aspects: The muon flux (40$\times$ lower than at surface) and angular distribution have been measured and matched by simulations  \cite{Ludwig2019}, as well as the neutron flux (180$\times$ lower than at surface) and energy spectrum \cite{Grieger2020}. A previous study of the high-energy, $E_\gamma >$ 3\,MeV, part of the $\gamma$-ray background also showed a strong reduction with respect to the surface \cite{Szuecs2019}. The present work complements these studies \cite{Ludwig2019,Grieger2020,Szuecs2019} by focusing on the low-energy, $E_\gamma \leq$ 3\,MeV,  part of the $\gamma$-ray background.

This work is organized as follows:
Section \ref{sec:Felsenkeller} describes the Felsenkeller shallow-underground laboratory, and section \ref{sec:Experiment} the detector and its passive and active shielding. The data analysis and results are shown in section \ref{sec:Analysis}. The sensitivity of this new setup for the example of $^7$Be is derived in section \ref{sec:7Be}. The data are discussed in section \ref{sec:Discussion}, and a conclusion and outlook are offered in section \ref{sec:Conclusion}.

\section{The Felsenkeller shallow-underground laboratory}
\label{sec:Felsenkeller}

\begin{figure}[t]
\centering
\includegraphics[width=\columnwidth,clip]{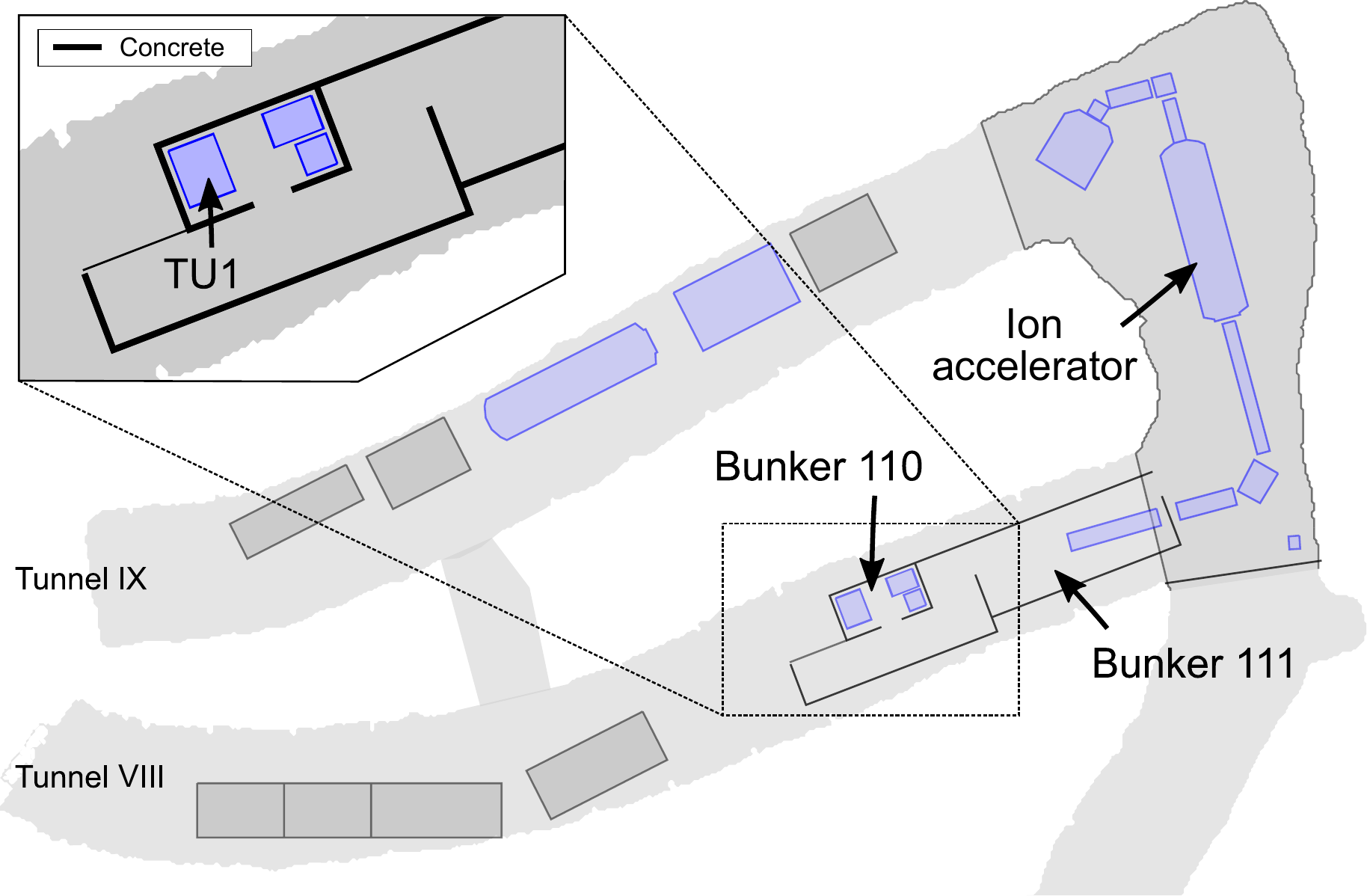}
\caption{Schematic layout of tunnels VIII and IX of the shallow-underground laboratory Felsenkeller in Dresden. The inlet shows bunker 110, which contains two coaxial HPGes (TU1 and TU4) and a well-type HPGe (TU2).}
\label{fig:FelsenkellerLayout}
\end{figure}

The Felsenkeller shallow-underground laboratory (Figure \ref{fig:FelsenkellerLayout}) is located in Dresden, Germany, and is shielded by a rock overburden of 45\,m (140\,m.w.e.) \cite{Ludwig2019}. It is part of a former industrial site and the laboratory itself is built into the tunnels VIII and IX of this area. The surrounding hornblende monzonite \cite{Paelchen2008} shows naturally containing $^{238}$U and $^{232}$Th with specific activities of 170(30)\,Bq/kg and 130(30)\,Bq/kg respectively \cite{Grieger2021}.

The laboratory hosts a 5\,MV Pelletron accelerator as well as two concrete-shielded bunkers (110 and 111). Bunker 110 is used for low-background offline measurements and is surrounded by 40\,cm of low-activity concrete (see section \ref{subsec:PassiveShielding} below for details). It currently hosts three HPGe detectors (called TU1, TU2 and TU4, respectively) for $\gamma$-ray spectrometry, as well as two silicon drift detectors (TU3 and TU5) for X-ray spectrometry. Bunker 111 hosts the target area of the accelerator and is used for in-beam $\gamma$-ray spectrometry measurements on radiative capture reactions.

In bunker 110, the angle-integrated muon flux density is $\varphi_\mu=5.4(4)$\,m$^{-2}$s$^{-1}$ corresponding to 140 meters of water equivalent (m.w.e.) \cite{Ludwig2019}. The neutron flux density is $\varphi_n=0.61(3)$\,m$^{-2}$s$^{-1}$ integrated over a broad energy interval ranging from 10$^{-9}$\,MeV to 300\,MeV \cite{Grieger2021, Grieger2020}.

\section{Experimental Setup}
\label{sec:Experiment}

\begin{figure*}[bt]
\centering
\includegraphics[width=\textwidth,trim=7.7mm 0mm 15mm 5mm,clip]{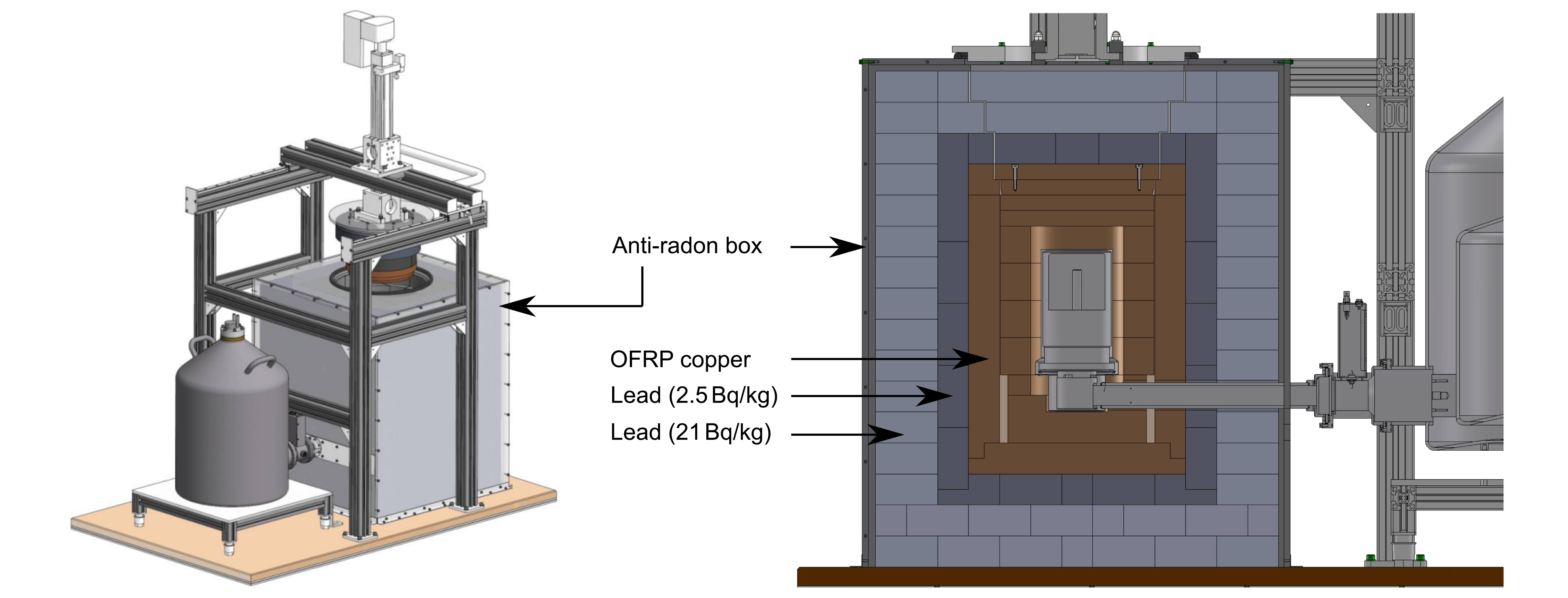}
\caption{Left panel: Schematic drawing of the passive shielding for the TU1 detector including the lifting mechanism for the lid and the anti-radon box. Right panel: Profile cut of the setup, from outside to inside: acrylic glass (anti-radon box), 21\,Bq/kg lead, 2.5\,Bq/kg lead, and OFRP copper. The muon veto panels (Figure \ref{fig:MuonVetoDrawing}) are placed just outside the anti-radon box but omitted here for clarity.}
\label{fig:TU1Drawing}
\end{figure*}

This work concentrates on the coaxial HPGe detector TU1. The other detectors will be described elsewhere.

\begin{figure}[b!!]
\centering
\includegraphics[width=0.8\columnwidth,clip]{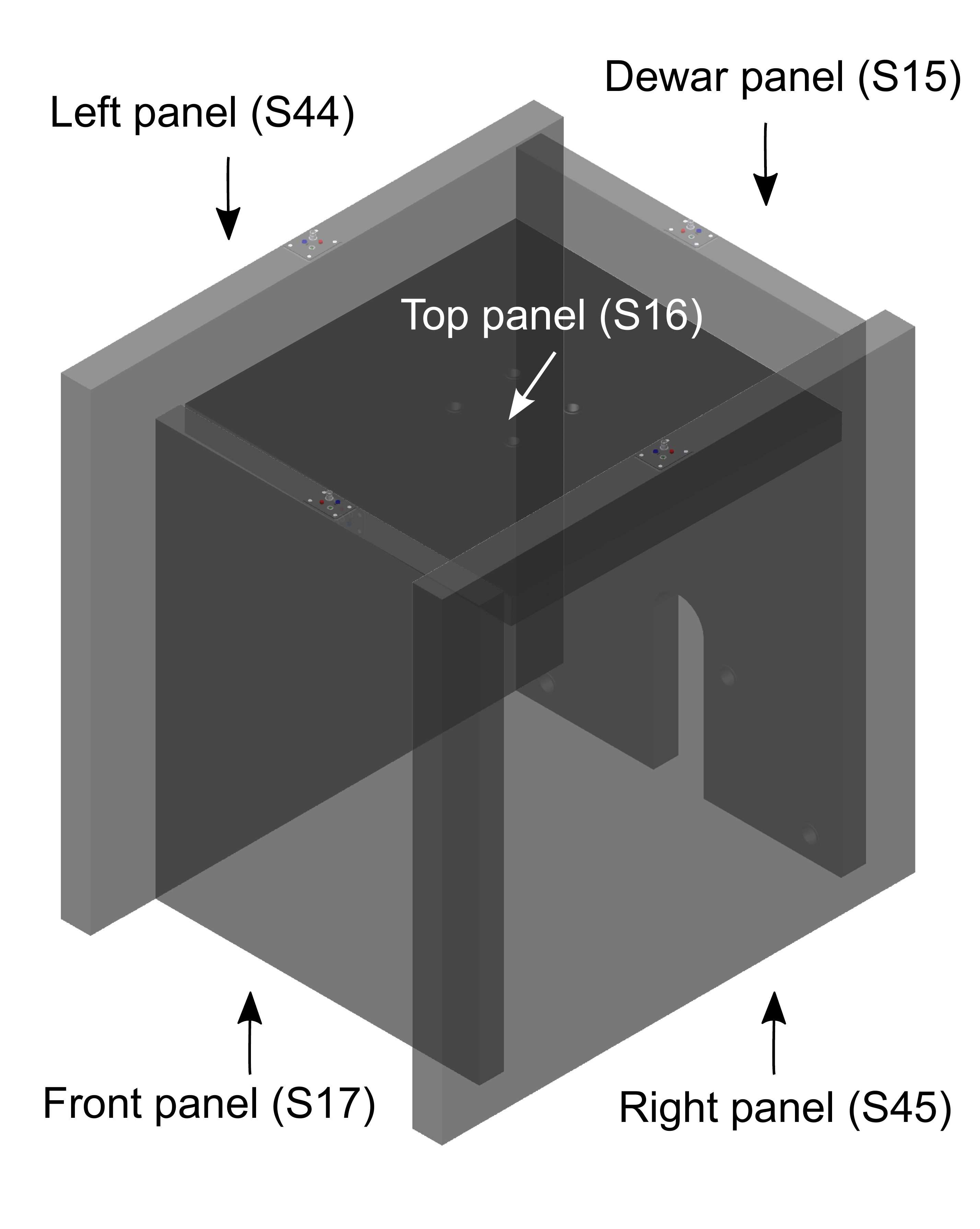}
\caption{Schematic drawing of the muon veto panels surrounding the anti-radon box (figure \ref{fig:TU1Drawing}). The top panel (S16) has four holes for the lifting mechanism, and the dewar-side panel (S15) has a recess for the cold finger and additional cutouts for nitrogen supply and overflow.}
\label{fig:MuonVetoDrawing}
\end{figure}

\subsection{The HPGe detector}
\label{hpgedetector}
The detector is a coaxial p-type high purity germanium detector (HPGe) with a relative efficiency of 163\,\%. It was produced by Mirion Technologies (Canberra) according to their ultra-low background (ULB) specifications and is of the type GX 150-250-R. The crystal has a mass of 3.06\,kg and a volume of 574\,cm$^3$ (length 90\,mm and diameter 90\,mm). An end cap of 1.6\,mm of aluminum and a polished-off p-layer of $<$0.5\,mm thickness enable measurements down to 22\,keV, well below the usual lower energy limit for standard p-type HPGe detectors.

A photon with 22\,keV induces a preamplified pulse height of approximately 6\,mV in this detector. The observed background noise is
$\leq 2.4$\,mVpp (95\,\% CL). The measured resolution at 1.333\,MeV $\gamma$-ray energy is 2.0 keV full width at half maximum (FWHM). There are no significant shoulders to the peak in the pulse height spectrum, and the ratio of full width at one fifth maximum (FWFM) and FWHM is 2.7.

\subsection{The passive shielding}
\label{subsec:PassiveShielding}	

The passive shielding of TU1 (Figure \ref{fig:TU1Drawing}) consists of several layers, which are listed here from the outside to the inside.
\begin{enumerate}
    \item The laboratory is surrounded in all directions by a 40\,cm thick layer of concrete. Prior to mixing the concrete, the solid components used had been studied individually by $\gamma$-ray spectrometry, and based on the known composition of the concrete, specific activities of 17(3) Bq/kg $^{238}$U and 18(2) Bq/kg $^{232}$Th were found, assuming secular equilibrium. For $^{40}$K, 280(30) Bq/kg was found. 
    \item The active veto consists of a 5\,cm thick layer of polyvinyltoluene (EJ-200\footnote{Eljen Technology, Sweetwater, Texas, USA.}) with a density of 1.0\,g/cm$^3$. It is listed here because it slightly attenuates incident $\gamma$-rays and moderates incident neutrons. The active veto is described below (section  \ref{subsec:MyonVeto}).
    \item A 1\,cm thick layer of acrylic glass forms an anti-radon box (section \ref{subsec:AntiRadonBox}).
    \item The outer lead layer has a thickness of 10\,cm and a specific $^{210}$Pb activity of 21(2)\,Bq/kg.
    \item  The inner lead layer is 5\,cm thick, with a specific $^{210}$Pb activity of 2.5(1)\,Bq/kg.
    \item The innermost layer of the passive shielding consists of 10\,cm of oxygen-free radio-pure (OFRP) copper ($\le 0.04$\,Bq/kg $^{238}$U). 
\end{enumerate}

For changing the sample, the lead castle can be opened using an electric lifting mechanism (Figure \ref{fig:TU1Drawing}). This mechanism lifts a cutout of the upper copper and lead bricks, as well as a part of the anti-radon box and the whole scintillation panel S16 (figure \ref{fig:MuonVetoDrawing}).

\subsection{The anti-radon box}
\label{subsec:AntiRadonBox}

The air in the bunkers of the Felsenkeller underground laboratory is exchanged four times per hour by an automatic ventilation system that is continuously active day and night. Part of the incoming air is fresh air brought in from the outside, and the remainder dried and recirculated. Prior to the construction of the laboratory and its mechanical ventilation, a radon concentration of 0-300\,Bq/m$^{3}$ was measured in tunnel VIII and IX, depending on weather conditions. After completion of the laboratory, and with the automatic ventilation system running, the radon concentration in bunker 110 has been remeasured. In 14\,days of measurements, it ranged from 0-53\,Bq/m$^3$, with an average of 11(7)\,Bq/m$^3$.

In order to minimize the impact \cite{Stekl2021, Koehler2009} of the remaining radon on the background rate of TU1, an airtight anti-radon box has been installed. The box is constructed from acrylic glass and flushed with radon-free nitrogen from the boil-off of the HPGe dewar, monitored by a bubbler. 

\subsection{The active muon veto}
\label{subsec:MyonVeto}

The active shielding for TU1 is composed of five large plastic (EJ200) scintillation panels from Scionix (figure \ref{fig:MuonVetoDrawing}) with 5\,cm thickness and a size ranging between 0.5\,m$^2$ (S16) and 1\,m$^2$ (S44 \& S45). The panels cover the anti-radon box from each side except from the bottom. EJ200 is a type of polyvinyltoluene with a light output of 10,000 \linebreak photons/MeV$_{ee}$ (MeV electron equivalent), a light attenuation length of 380\,cm, a rise time of 0.9\,ns and a decay time constant of 2.1\,ns, making it suitable for timing measurements in the 0.1\,ns range even in scintillating detectors as large as several meters \cite{Reinhardt16-NIMA}.

The scintillation light is read out by ET9900 photomultiplier tubes\footnote{ET Enterprises, Ltd., Uxbridge, UK.} (PMTs) that are integrated within the scintillation panels and that have a 2$\pi$ sensitive solid angle. Also the high voltage supplies of the PMTs are included in the panel, so that only two connections are required on one side of each panel: an input for low-voltage power (LEMO 00), and the output for the PMT anode signal (BNC). The scintillating panels are coated with a reflector and wrapped in lightproof vinyl. 

\begin{figure*}[t]
\centering
\includegraphics[width=\textwidth,trim=7.7mm 0mm 15mm 5mm,clip]{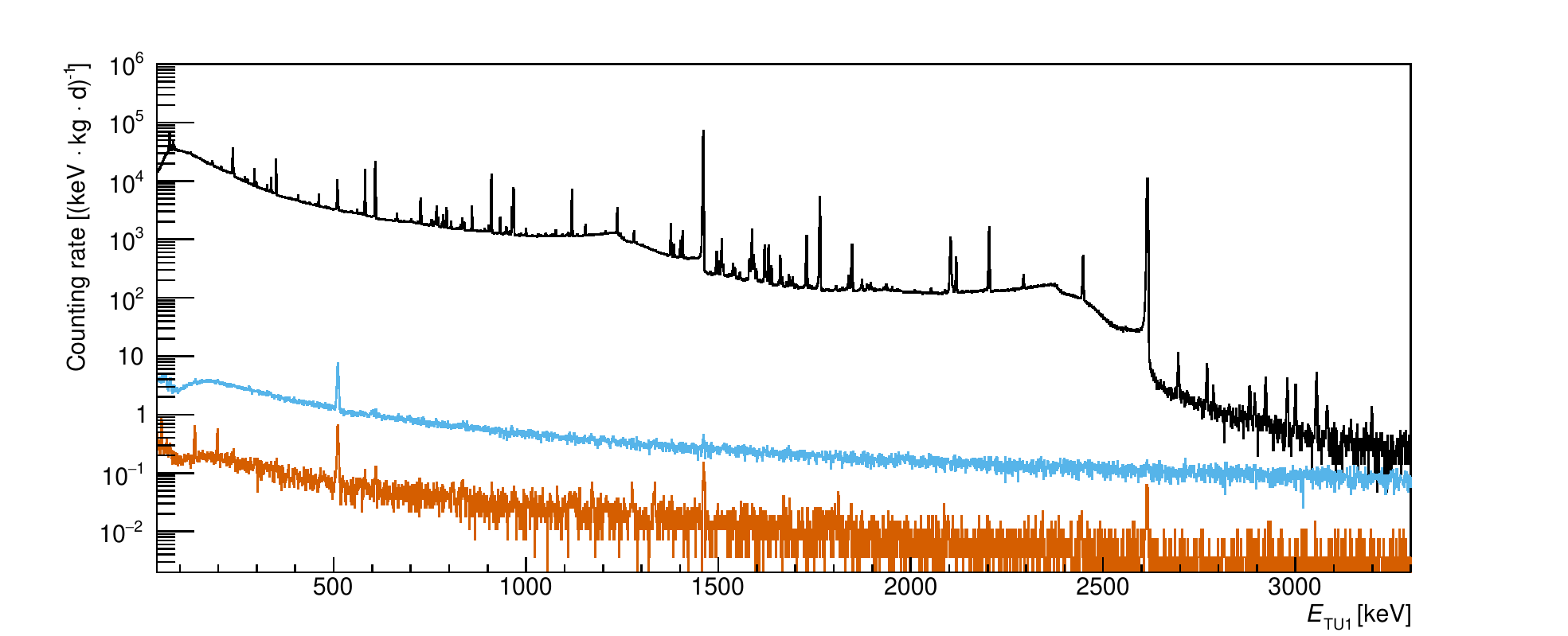}
\caption{Pulse height spectrum for the TU1 detector, normalized to running time, bin width, and detector mass: Bunker 110, without any shielding except for the concrete walls (black curve, 6.8\,days running time). Full passive shielding and anti-radon box (blue curve, 90.0\,days). The same spectrum, but applying also the active shield (red curve). The $\gamma$-lines identified in the red spectrum are listed in Table \ref{Table:Gammalines}. See text for details.}
\label{fig:TU1_spectrum}
\end{figure*}

\subsection{Electronics and data acquisition}
\label{subsec:DAQ}

The signals from the six detectors (TU1 and five veto panels) are recorded in list mode using a CAEN DT5725S digitizer. This module includes eight channels with 14\,bit resolution at 250\,MS/s sampling rate. All channels share the same internal clock, but each channel is separately triggered by the digital trigger included in the DT5725S device. The recorded waveforms are converted to a time stamp and pulse height with a trapezoidal filter by the CAEN pulse height firmware, version DPP-PHA 4.22\_139.130, directly inside the DT5725S unit. The event-by-event time stamp, pulse height, and flags further characterizing the event as possible pile-up and overflow are saved in a buffer, transferred via USB cable to the computer and saved on hard disk. 

For the scintillating panels, typical trigger rates of 9-140\,s$^{-1}$ are observed per panel. For the TU1 detector, the typical trigger rate without sample is 5\,s$^{-1}$. Given the typical time lengths of the trapezoidal filter of 4.4\,$\upmu$s (TU1) and 5.0\,$\upmu$s (scintillators), dead time and pile-up effects are expected to be insignificant. The data analysis is performed offline.


\section{Data analysis}
\label{sec:Analysis}

In this section, the offline data analysis is described, including the optimization of the active veto parameters.

\subsection{Pulse height spectrum in TU1 using only the passive shield}
\label{subsec:TU1_onlyPassiveShield}

The pulse height spectrum of the TU1 detector, measured within bunker 110 without any additional shielding, displays many $\gamma$ lines due to the natural decay chains (Figure \ref{fig:TU1_spectrum}, black spectrum). The $^{40}$K and $^{208}$Tl lines emerge prominently. In addition, there are a number of neutron-induced features due to the remaining neutron background \cite{Grieger2020} that can be picked out due to their triangular shapes and above 2.6\,MeV, there are a number of weak branches from $^{208}$Tl. The $\gamma$-ray energy resolution at the $^{40}$K peak (1460.8\,keV) is 3.1\,keV (full width at half maximum, FWHM) in the adopted digitizer-based DAQ scheme, worse than the 2.2\,keV FWHM measured with the same detector and analog electronics.

When applying the full passive shielding, the continuum below 2.6\,MeV is reduced by three to five orders of magnitude, and the neutron features are no longer apparent (Figure \ref{fig:TU1_spectrum}, blue spectrum). Instead, a wide continuum appears that is only slightly energy dependent, showing the effects of the energy loss of the remaining cosmic-ray induced muon flux.

\begin{figure}[bt]
\centering
\includegraphics[width=\columnwidth,trim=0mm 0mm 5mm 15mm,clip]{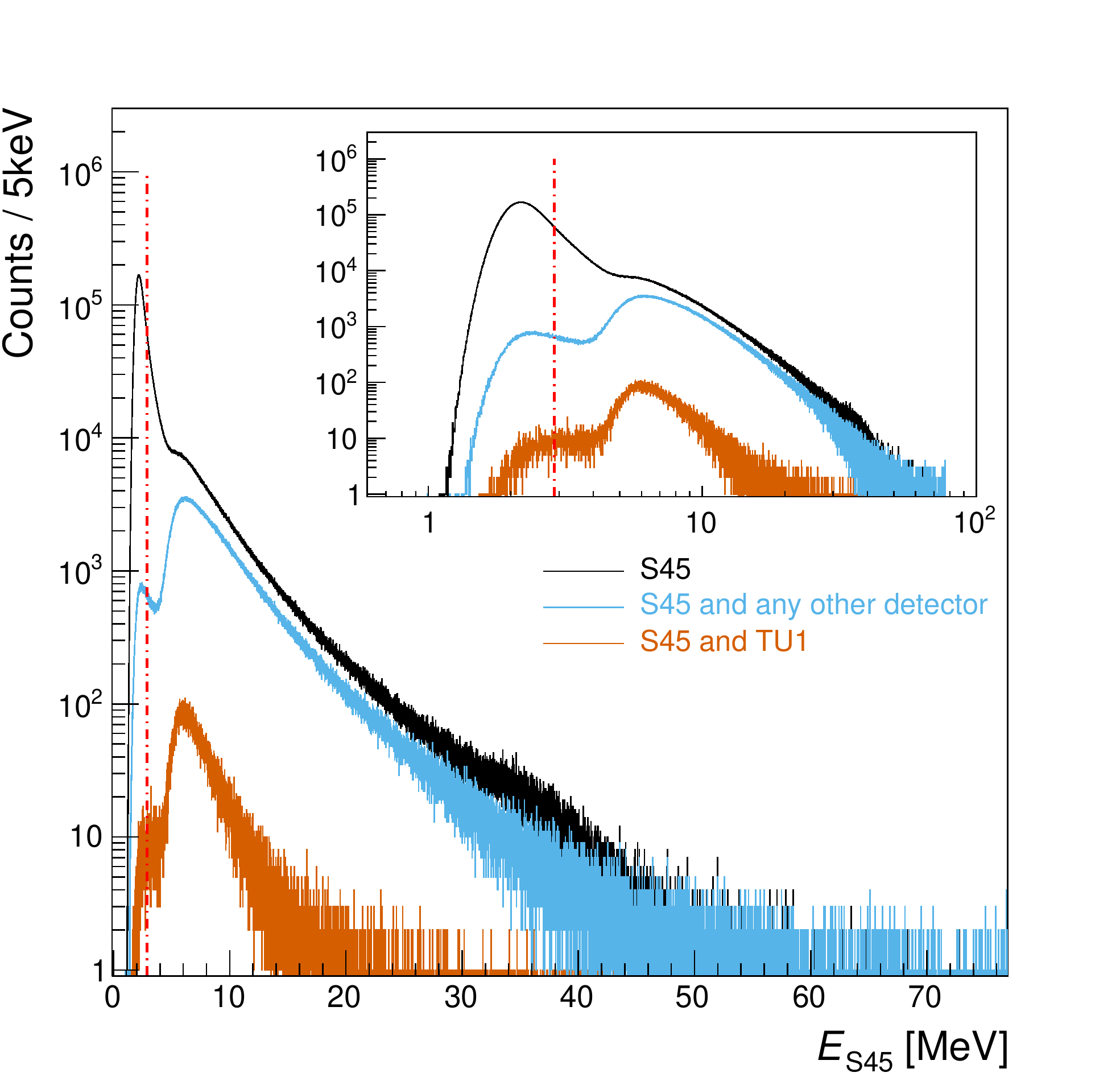}
\caption{Pulse height spectrum of scintillation panel S45, 29.5\,days running time. Black curve: raw spectrum. Blue curve: requiring coincidence within [-5\,$\upmu$s,+5\,$\upmu$s] with any of \{S15, S16, S17, S44\} or TU1 ($E_{\rm TU1} \geq 0.04$\,MeV). Red curve: requiring coincidence with TU1
($E_{\rm TU1} \geq 0.04$\,MeV). Red dashed line: energy cut for this panel (2.9\,MeV, section \ref{subsec:EnergyCut}). See text for details.
}
\label{fig:SpectrumScint45}
\end{figure}

\subsection{Active muon veto}
\label{subsec:AnalysisActiveMuonVeto}

In a first step, the pulse height spectrum of the scintillating panels has been calibrated in energy using the Compton edges of radionuclide standards made of $^{137}$Cs, $^{60}$Co, and $^{88}$Y. These point-like standards were placed on the center of one of the large sides of each scintillating panel, in turn. As a second step, the trigger threshold for the scintillating panels was increased to approximately 2\,MeV, in order to avoid unnecessarily recording radionuclide-induced pulses detected in the scintillators.

The resulting energy-calibrated spectrum (Figure \ref{fig:SpectrumScint45}, black curve) shows a peak near 2\,MeV$_{ee}$ that is given by the above mentioned trigger condition and a broad peak near 6\,MeV$_{ee}$ that is assumed to be due to muons. 

The observed pulse height in the histogram is proportional to the number of scintillation photons detected by the PMT. Due to the finite reflectivity of the reflective wrapping, as well as the light attenuation in the crystal itself, not all of the emitted photons are detected. This is even enhanced for panels S15 and S16 which also include holes and a recess. The effect has been studied using a $^{60}$Co radionuclide standard placed at several places across the scintillating panel. Using the light yield at the center of the panel as reference, $>$90\% of the area of the panel shows observed pulse heights between 0.7 and 1.2 times the reference pulse height. The remaining $<$10\% are located very close to the PMT and show higher light yields. It is up to 5 times the reference light yield when the $^{60}$Co source is placed directly atop the sensitive area of the PMT, so that it may see scintillation light from both sides. 

At the depth of the Felsenkeller lab, the muon energy spectrum peaks at approximately 30\,GeV, and the energy loss of muons is about 2\,MeV$_{ee}$ per cm of EJ-200 material traversed \cite{Groom2001}. Qualitatively, the large cross section of the panels is expected to lead to muons going perpendicularly through the panel dominating the spectrum. They would lose approximately 10\,MeV$_{ee}$, and applying the above mentioned factor of 0.7 for less than ideal light collection, would register near 7\,MeV$_{ee}$, consistent with the observed peak at 6\,MeV$_{ee}$.

In addition, many muon paths are possible that are significantly longer than 5\,cm. As a consequence, the muon-induced spectrum in S45 extends to rather high energy, up to dozens of MeV$_{ee}$ (Figure \ref{fig:SpectrumScint45}, black curve).

In the following two subsections, by offline analysis of background runs lasting several days, the two main parameters for the active muon veto are optimized: First, the timing coincidence condition (section \ref{subsec:TimeCoincidence}), and second, the pulse height threshold (section \ref{subsec:EnergyCut}). A third subsection (\ref{subsec:TimeEnergy}) then discusses additional information gained by combining the two cuts.

\subsection{Time coincidence condition}
\label{subsec:TimeCoincidence}

\begin{figure}[bt]
\centering
\includegraphics[width=\columnwidth,trim=0mm 0mm 5mm 15mm,clip]{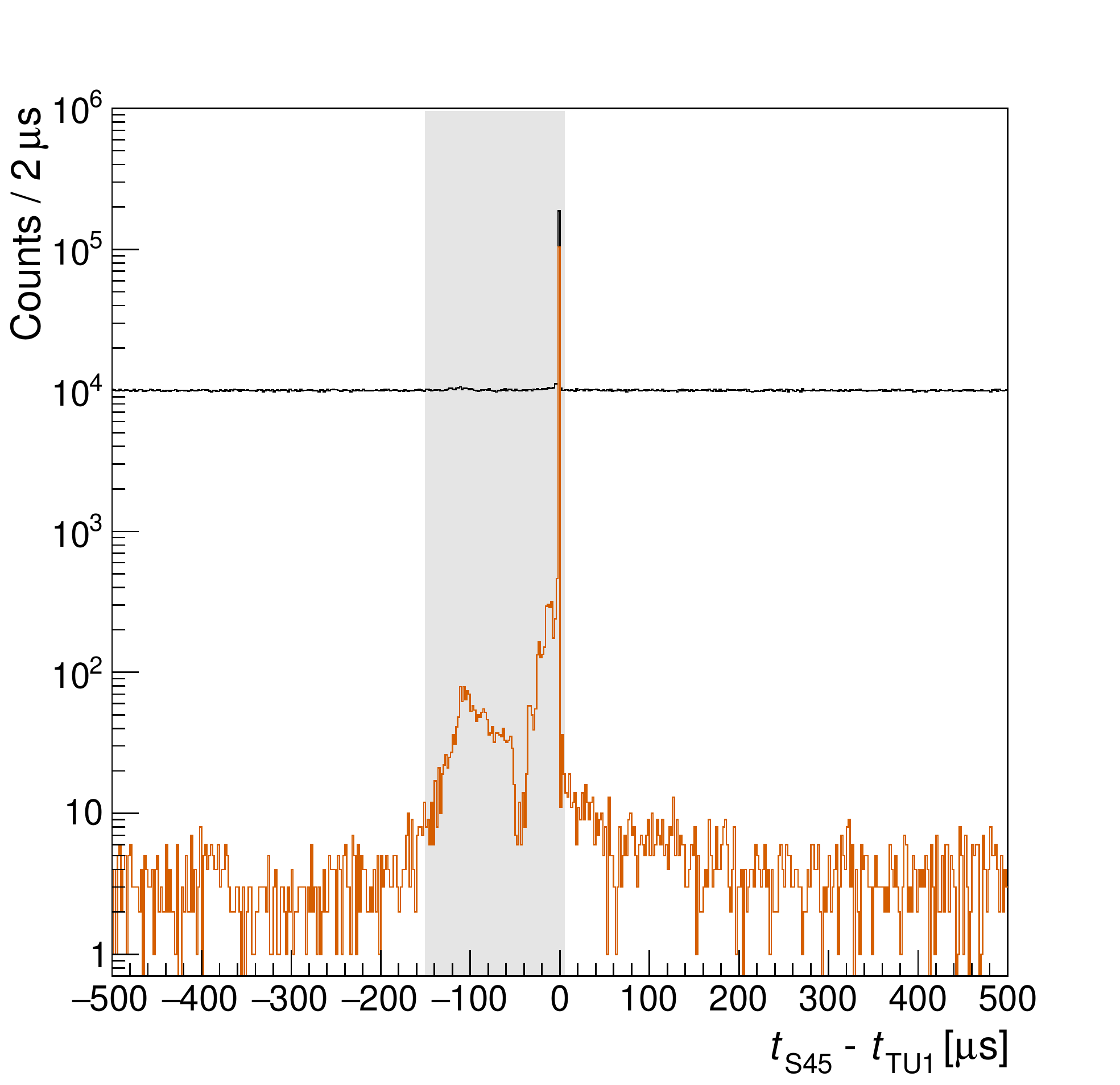}
\caption{Black curve: Event by event time difference between events in scintillation panel S45 ($t_{\rm S45}$) and in TU1 ($t_{\rm TU1}$), 29.5\,days. Red curve, in addition $E_{\rm TU1}\geq$ 0.04\,MeV and $E_{\rm S45}\geq E_\textrm{thresh, S45}$ ($E_\textrm{thresh, S45}$ = 2.9\,MeV) are required. The grey box shows the ``long" coincidence timing window of [-150\,$\upmu$s,+5\,$\upmu$s]. See text for details.}
\label{fig:TimingHistogram}
\end{figure}

The time difference between the trigger times $t_{\rm TU1}$ in TU1 and those of the scintillating panels $t_{i}$ (with $i \in \{{\rm S15,S16,S17,S44,S45}\}$) shows a sharp peak that is contained in the time interval [-5\,$\upmu$s,+5\,$\upmu$s]. A typical example for this time difference spectrum is given in Figure \ref{fig:TimingHistogram}. The mutual time differences between individual scintillator panels are not shown but display a similar pattern. 

The sharp peak in the [-5\,$\upmu$s,+5\,$\upmu$s] coincidence window of figure \ref{fig:TimingHistogram} is caused by muons that pass both S45 and TU1. The peak width is given by the time resolution of TU1, which is approximately on the level of the 2\,$\upmu$s binning used in Figure \ref{fig:TimingHistogram}. A second, longer time coincidence window of [-150\,$\upmu$s,+5\,$\upmu$s]  will be introduced below.

\subsection{Energy threshold in the scintillator panels}
\label{subsec:EnergyCut}

Even with a trigger threshold near 2\,MeV (section \ref{subsec:AnalysisActiveMuonVeto}), there is still a significant number of radionuclide-induced events in the pulse height spectrum of the scintillator. This can be seen by comparing the free-running pulse height spectrum of one scintillator panel  (Figure \ref{fig:SpectrumScint45}, black curve) with the same spectrum when requiring time coincidence, within a ``short" coincidence window of [-5\,$\upmu$s,+5\,$\upmu$s], with any of the other detectors  (Figure \ref{fig:SpectrumScint45}, blue curve).

In order to reduce the impact of these remaining radio\-nuclide-induced events, an additional threshold energy \linebreak $E_{{\rm thresh}, i}$  (with $i \in \{{\rm S15,S16,S17,S44,S45}\}$) is defined. This threshold was carefully optimized. Too low  $E_{{\rm thresh}, i}$ would allow a too large impact of radionuclide-induced events, whereas a too high  $E_{{\rm thresh}, i}$ would reduce the muon veto efficiency. 

This optimization was carried by analyzing a one day long run with a 2\,kBq $^7$Be source mounted in close geometry. Starting from the local minimum to the left of the muon peak ($ E_{{\rm min, \, S45}}$ = 3.6\,MeV for the blue curve in Figure \ref{fig:SpectrumScint45}), the energy threshold was varied in the offline analysis. Then, for each given value of the energy threshold, two quantities were determined. First, the cut survival probability $P_{\rm cut \, survival}$ for events in the $^7$Be peak area in the TU1 detector given by
\begin{equation}
    P_{\rm cut \, survival} = \frac{N_{\rm TU1}(^7{\rm Be})|_{\rm cut}}{N_{\rm TU1}(^7{\rm Be})|_{\rm free \, running}} 
\end{equation}
The second quantity was the no-source specific background rate $\dot{N}_{40-2700}$ in the 40-2700\,keV energy region. It is determined by analyzing a 30-day run without any source and with the same running conditions as the $^7$Be run with the same energy threshold values and requiring anticoincidence with all the scintillator panels within a time window [-150\,$\upmu$s;+5\,$\upmu$s]. See section \ref{subsec:TimeEnergy} below for the justification for this somewhat enlarged time coincidence window. 

The optimal energy threshold was then defined to be the energy for which $P_{\rm cut \, survival}$ = 0.995 was found. This was typically at 0.8$\times$ the local minimum energy (red dashed line at $E_{\rm thresh, \, S45}$ = 2.9\,MeV in Figure \ref{fig:SpectrumScint45}). For the other scintillators, not shown in the Figure, similar energy thresholds of 2-3\,MeV were determined. 

The impact of the choice of the energy threshold is illustrated in Table \ref{Table:Sensitivity} where all energy thresholds are varied together by a constant factor $f = E_{{\rm thresh}, i} / E_{{\rm min}, i}$, and for each value of $f$ the cut survival probability and background count rate are shown.

\begin{table}
\caption{%
\label{Table:Ethresh}%
Impact of the variation of the energy threshold by a factor $f$ on the cut survival probability $P_{\rm cut \, survival}$ and on the specific background rate $\dot{N}_{40-2700}$. See text for details.
} 
\begin{tabular}{cccl
		}
\hline 
$f$ & $P_{\rm cut \, survival}$ & $\dot{N}_{40-2700}$ & Remark\\ \hline
 0.4 & 0.958(1) & 112(1) \\ 
 0.6 & 0.980(1) & 113(1) \\
 0.8 & 0.995(1) & 116(1) & Adopted value\\
 1.0 & 0.998(1) & 120(1)\\
 1.2 & 0.998(1) & 124(1)\\
\hline
\end{tabular}
\end{table}

It is clear from Table \ref{Table:Ethresh} that the background counting rate changes by only a few percent in the region studied, while the cut survival probability is rather sensitive to the choice of the energy threshold. A value of  $P_{\rm cut \, survival}$ =0.995 was adopted. This value introduces a small systematic shift, no more than -0.5\%, which is lower than the typical geometry-dependent detection efficiency uncertainty of 1-3\%. This adopted value does not significantly worsen the precision of the activity determination.

As a final check for the adopted threshold energy, the ratio of counts in the S45 pulse height spectrum requiring coincidence with TU1 (red curve in Figure \ref{fig:SpectrumScint45}) with the S45 free running spectrum (black curve in Figure \ref{fig:SpectrumScint45}) is computed. For $E_{\rm S45} < E_{\rm thresh, \, S45}$, the observed ratio is 4.9$\times$10$^{-5}$. This is consistent with expectation for random coincidences, given the 10\,$\upmu$s width of the ``short" time coincidence window and the TU1 trigger rate of 4.9\,s$^{-1}$. For $E_{\rm S45} > E_{\rm thresh, \, S45}$, instead, the observed ratio is much larger, 4.3$\times$10$^{-3}$. In conclusion, S45-TU1 coincident events can be fully explained by random coincidences for 
 $E_{\rm S45} < E_{\rm thresh, \, S45}$, while only 1\% of S45-TU1 coincident events with  $E_{\rm S45} > E_{\rm thresh, \, S45}$ can be explained by random coincidences. Thus, it also confirms that the threshold value $E_{\rm thresh, \, S45}$ is appropriately chosen.
 
\subsection{Combination of timing and energy cuts}
\label{subsec:TimeEnergy}

As a next step, the newly determined energy threshold is adopted, and the time difference spectrum between TU1 and S45 is re-determined, now requiring $E_{\rm S45}\geq E_\textrm{thresh, S45}$ ($E_\textrm{thresh, S45}$ and  $E_{\rm TU1}\geq 0.04$\,MeV). As expected, the resultant spectrum (Figure \ref{fig:TimingHistogram}, red curve) still shows the prompt muon coincidence peak at -5 $\upmu$s $\leq (t_{\rm S45}-t_{\rm TU1}) \leq$ 5$\upmu$s. 

In addition, now two distinct features emerge that had previously been covered by the radionuclide-induced events, namely at time differences $(t_{\rm S45}-t_{\rm TU1}) \approx$ -100\,$\upmu$s and $(t_{\rm S45}-t_{\rm TU1}) \approx$ -10\,$\upmu$s, respectively. 

The first feature may be due to muons which first pass the scintillator panel and subsequently scatter on the lead shield to create a free neutron. This ($\upmu$,n) process has previously been found to dominate the neutron flux, especially at high neutron energies, in the Felsenkeller laboratory \cite{Grieger2020}. The delay of $\sim$100\,$\upmu$s between the passage of the muon through the scintillator panel and the signal in the HPGe detector may be due to free ($\upmu$,n) neutrons that, after creation, are first scattered a few times, and slightly moderated, in the lead shield before they give rise to a signal in the HPGe detector. 

The second feature, with a delay of typically 10\,$\upmu$s or less between the passage of the muon through the scintillator, may be due to the decay of stopped muons (mean lifetime 2.2\,$\upmu$s) in the lead shield, and subsequent detection of secondary electrons, positrons, or annihilation radiation from the decay products.

The specific rate of events within the extended part of the anticoincidence window, i.e. 
[-150\,$\upmu$s;-5\,$\upmu$s], that also satisfies the energy cuts for TU1 and the scintillators is 20(1)\,kg$^{-1}$d$^{-1}$. Thus, the no-source background rate (Table \ref{Table:Ethresh}) of 116\,kg$^{-1}$d$^{-1}$ for the full anticoincidence window of [-150\,$\upmu$s;+5\,$\upmu$s] would increase to 136\,kg$^{-1}$d$^{-1}$, if a more narrow anticoincidence window of [-5\,$\upmu$s;+5\,$\upmu$s] would be used.

\subsection{Rates of coincident events}
\label{RatesOfCoincidences}

The coincidence rates ($E_{\textrm{TU1}}\geq 40\,$keV, $E_\textrm{panel}\geq E_\textrm{thresh}$, time coincidence window $t_\textrm{cut}$=[-150$\,\upmu$s,+5\,$\upmu$s]) between the scintillation panels and TU1 are $R=0.0207(1)\textrm{s}^{-1}$ (TU1$\wedge$S15), $0.0337(1)\textrm{s}^{-1}$ (TU1$\wedge$S16), $R=0.0208(1)\textrm{s}^{-1}$ (TU1$\wedge$S17), $0.0298(1)\textrm{s}^{-1}$ (TU1$\wedge$S44), and $0.0273(1)\textrm{s}^{-1}$ (TU1$\wedge$S45), respectively.

As expected, the coincidence ratios for the panels on opposite sides are close: The opposing panels S44 and S45 differ by just 9\%, and the opposing panels S15 and S17 by $<$1\%. The highest absolute coincidence rate is found for TU1$\wedge$S16, consistent with expectation based on the measured muon angular distribution in this area \cite{Ludwig2019}. 

The total rate of vetoed events in TU1 is $0.0763(2)\textrm{s}^{-1}$ for $E_{\textrm{TU1}}\geq0.04\,$MeV and the ``long" anticoincidence window of [-150$\,\upmu$s,+5\,$\upmu$s]. This is consistent with the expected rate of muons penetrating TU1 of 0.08\,s$^{-1}$. This rate has been estimated using the muon flux of  $\varphi_\upmu=5.4(4)\textrm{m}^{-2}\textrm{s}^{-1}$ that has previously been measured at this location \cite{Ludwig2019} and an estimated muon-exposed cross section of TU1 of 145\,cm$^2$.

\subsection{Preamplified signals of opposite polarity}
\label{subsec:OppositePolarity}

An uncommon characteristic of the TU1 detector is the fact that there is a significant rate, 6\,s$^{-1}$, of signals with unphysical polarity, opposite to the usual one,  at the preamplifier output. These pulses have the same general shape and amplitude as the correctly preamplified pulses, but opposite polarity. However, they are not correlated in time to any other events, either from TU1 or from any of the scintillating panels. For safety, these unphysical events are studied via a dedicated channel of the data acquisition (self-triggered just as the other channels) and stored for possible future analysis. The observed pulse height spectrum does not show the characteristic features of an HPGe spectrum, but rather resembles a Landau distribution. 

The opposite polarity events are believed to be caused by an electronic noise source in the preamplifier unit. Given their rate and random distribution in time, there is a 6$\times$10$^{-6}$ probability for them to affect a true physical event, which is negligible in the present, low-background application. 

\subsection{Dead time}
\label{subsec:Deadtime}

The total dead time $t_\textrm{d,tot}$ of TU1 is given by the dead time $t_\textrm{d,raw}$ due to the $\sim$10\,$\upmu$s processing time of the digitizer and an additional cut efficiency $\varepsilon_\textrm{cut}$ (section \ref{subsec:EnergyCut}), which takes into account the loss of counts due to random anticoincidences.

The digitizer dead time has been estimated based on the count rate of the resulting spectrum (output count rate, OCR) and the 1\,$\upmu$s trigger hold-off time during which the digitizer unit does not accept new triggers. In a procedure similar to the one recommended by CAEN \cite{CAEN}, the unknown input count rate (ICR) is determined in an iterative procedure. From a starting ICR and assuming Poisson-distributed events, a predicted OCR is determined and then compared to the observed OCR in order to obtain an improved ICR estimate for the next iteration. This iteration converges typically after five cycles. 

The TU1 channel of the digitizer self-triggers on the absolute height of the second derivative of the pulse, which triggers both for negative (physical) and positive (unphysical) pulses. In the TU1 channel, the unphysical events are converted to channel 0 of the pulse height histogram and fully taken into account during the dead time determination. They contribute a relative dead time of 6$\times$10$^{-6}$. In addition, the unphysical events are converted and stored in a separate channel of the digitizer, see section \ref{subsec:OppositePolarity} above. 

In a typical run, about 2\% of the detected events in the TU1 channel do not follow a Poisson distribution in time, but instead follow directly a preceding event. This behaviour is consistent with electronic noise or mechanical vibrations. Due to their limited number, these non-Poissonian events do not have a significant impact on the precision of the dead time determination.

\begin{table}[tb]
\caption{%
\label{Table:Gammalines}%
Energy and counting rates of the remaining $\gamma$-lines in the  TU1 detector, after applying the active veto. See text for details.} 
\begin{tabular}{S[
		table-number-alignment = right,
		separate-uncertainty = false,
		table-figures-uncertainty = 0,
		table-figures-decimal = 1,
		table-figures-integer = 1]
		cl
		S[
		table-number-alignment = left,
		separate-uncertainty = false,
		table-figures-uncertainty = 0,
		table-figures-decimal = 1,
		table-figures-integer = 1]
		}
\hline 
\multicolumn{1}{c}{Energy} & \multicolumn{1}{c}{Nuclide} & \multicolumn{1}{c}{Origin} & \multicolumn{1}{c}{Rate} \\ 
\multicolumn{1}{c}{[keV]} & & &
\multicolumn{1}{c}{[kg$^{-1}$d$^{-1}$]} \\ \hline
\multicolumn{1}{c}{52.0-55.6} & $^{73\textrm{m}}$Ge & $^{72}$Ge($n$,$\gamma$), $^{74}$Ge($n$,2$n$) & 1.28(19) \\
\multicolumn{1}{c}{63.9-66.7} & $^{73\textrm{m}}$Ge & $^{72}$Ge($n$,$\gamma$), $^{74}$Ge($n$,2$n$) & 0.38(17) \\
\multicolumn{1}{c}{72.8-75.0} & Pb\,X & Lead  $K_{\alpha}$\,\&\,$K_{\beta}$ X-rays & 0.33(12) \\
139.7 & $^{75\textrm{m}}$Ge& $^{74}$Ge($n$,$\gamma$), $^{76}$Ge($n$,2$n$) & 1.34(16) \\
198.4 & $^{71\textrm{m}}$Ge& $^{70}$Ge($n$,$\gamma$), $^{72}$Ge($n$,2$n$) & 1.00(15) \\
351.9 & $^{214}$Pb & $^{238}$U decay chain & 0.29(11) \\
511 & $e^+e^-$& Annihilation & 2.64(13) \\
583.2 & $^{208}$Tl & $^{232}$Th decay chain & 0.14(8) \\
609.3 & $^{214}$Bi & $^{238}$U decay chain & 0.21(8) \\
834.8 & $^{54}$Mn& $^{54}$Fe($n,p$)$^{54}$Mn & 0.17(7) \\
1173.2 & $^{60}$Co & $^{63}$Cu($n,\alpha$)$^{60}$Co & 0.18(5) \\
1274.5 & $^{22}$Na & Lab-specific nuclide & 0.20(5) \\
1332.5 & $^{60}$Co & $^{63}$Cu($n,\alpha$)$^{60}$Co & 0.16(5) \\
1460.8 & $^{40}$K & Natural radioactivity & 0.49(6) \\
2614.5 & $^{208}$Tl & $^{232}$Th decay chain & 0.22(4) \\ \hline
\end{tabular}
\end{table}

\subsection{Remaining background in TU1}

The remaining background after the construction of the passive shielding and the application of the active veto is shown as an orange histogram in figure \ref{fig:TU1_spectrum}. The remaining $\gamma$-lines, as well as their origins and counting rates, are listed in table \ref{Table:Gammalines}, and will be discussed in the following.

\subsubsection{Neutron induced lines in germanium}
\label{subsec:NeutronInducedPrompt}

The four low energetic $\gamma$-lines at 52-56, 64-67, 140, and 198\,keV are due to isomeric states in several germanium isotopes. These states in $^{71\textrm{m}}$Ge ($E_x=198.4\,$keV, 20\,ms half-life), $^{73\textrm{m}}$Ge (66.7\,keV, 0.5\,s half-life), and $^{75\textrm{m}}$Ge (139.7\,keV, 48\,s half-life) are populated by neutron captures, ($n,2n$) reactions, or ($n,n'\gamma$) reactions  within the germanium crystal itself. All three have half-lives that are too long for an effective active veto.

The peaks at 52-56 and 64-67\,keV (Table \ref{Table:Gammalines}) are due to the decay of the $E_{x}$ = 67\,keV metastable second excited state of $^{73}$Ge \cite{Bunting74-NIM}. 
The half-life of the first excited state ($T_{1/2}=2.9\,\upmu$s) is comparable to the length of the timing window for the HPGe, giving rise to partial summation effects. If the decay of the first excited state takes enough time, the lower-energy cluster of peaks in the TU1 spectrum will be fed by the 53.4\,keV photon (or the 52.0\,keV-53.4\,keV conversion electron respectively), which can be separated by the software from the subsequent decay. If this decay takes a shorter amount of time though, both peak clusters in the TU1 spectrum can be fed due to summation effects. The lower-energy one via the 53.4\,keV photon (or the 52.0\,keV-53.4\,keV conversion electron respectively) and their summation with the subsequently emitted conversion electron of 2.2\,keV. The higher-energy cluster then originates from the summation of the 53.4\,keV photon (or the 52.0\,keV-53.4\,keV conversion electron respectively) and their subsequently emitted conversion electron with 11.9\,keV-13.3\,keV. 

The 139.7\,keV $\gamma$ ray is due to the decay of the $^{75\textrm{m}}$Ge state at the same energy to the ground state of $^{75}$Ge. 
The rate of this $\gamma$ line can be used to approximate the thermal neutron flux density inside the passive shielding, resulting in $\varphi=(3.2 \pm 1.0)\times 10^{-5}\,\textrm{cm}^{-2}\textrm{s}^{-1}$ using a conversion factor from literature \cite{Skoro1992}. During a dedicated campaign to measure the neutron flux in the Felsenkeller laboratory \cite{Grieger2020}, the thermal flux inside room 110 has been measured. The thermal flux obtained with an unmoderated $^3$He tube during a run previous to the installation of TU1, i.e. in an empty bunker 110, with its lead shield is ($2.3 \pm 0.2) \times 10^{-5}\,\textrm{cm}^{-2}\textrm{s}^{-1}$ \cite{Grieger2021,Hensel19-Master}. It is known that large passive lead shields can enhance the neutron flux in a shallow-underground laboratory such as Felsenkeller \cite{Grieger2021}. This is mainly caused by ($\mu,n$) reactions in the high atomic charge shielding material. Neutrons of $\sim$ MeV kinetic energy are produced in these reactions \cite{Grieger2020,Grieger2021} and subsequently moderated, for example, in the 200\,kg of EJ-200 organic scintillator forming the muon veto (section \ref{subsec:MyonVeto}).

Finally, the 198.4\,keV peak is caused by the decay of the metastable level of $^{71}$Ge \cite{Bunting74-NIM,Heusser93-NIMB,Anders2013}, which has a half-life of 20\,ms. Its decay scheme suggests a chain emission of, first, a 12-22\,keV conversion electron and, second a $gamma$-ray of $E=175.0\,$keV \cite{Abusaleem2011}. The peak in the spectrum is a summation line, which benefits from both a negligible escaping probability of the emitted conversion electrons and the short half-life of the first excited state of $^{71}$Ge at $E_x=175\,$keV (T$_{1/2}=79$\,ns) with respect to the comparatively longer timing window of the HPGe (section \ref{subsec:TimeCoincidence}).

\subsubsection{Neutron activation lines}
\label{subsec:NeutronInducedActivation}

The spectrum shows three lines from the decay of long-lived radionuclides that may be attributed to neutron activation: $^{54}$Mn ($E_\gamma$ = 834.8\,keV, $T_{1/2}=312$\,d) may be produced via the $^{54}$Fe($n$,$p$)$^{54}$Mn reaction, and $^{60}$Co (1173.2 and 1332.5\,keV, $T_{1/2}=1925$\,d) via the $^{63}$Cu($n,\alpha$)$^{60}$Co reaction. These nuclides are typical contaminations in low-background HPGe detectors. They are likely due to cosmic-ray activation on shielding or structural materials.

It is expected that these contaminations slowly decay out in the next years and stabilize at a much lower level, consistent with the neutron flux in Felsenkeller that is 180 times lower than at surface \cite{Grieger2020}.

\subsubsection{Annihilation peak at 511\,keV}

 The annihilation peak at 511\,keV is attenuated by a factor of 11 by the active veto. The veto thus removes most of the muon-induced effects, including the decay of stopped muons. 
 
The remaining 511 keV counting rate is largely due to the annihilation of positrons emitted in $\beta^+$ decays in the detector and shielding material. It is noted that there is a small contribution ($\sim$0.04\,kg$^{-1}$d$^{-1}$) due to photons with $E_\gamma$ = 510.77\,keV from the decay of $^{208}$Tl, which also causes the 583 and 2615\,keV lines. 
 
\subsubsection{Radon-induced effects}

The low specific radon activity of 11(7) Bq/m$^3$ in the measurement bunker (section \ref{subsec:AntiRadonBox}) may lead to small amounts of radioactive $^{222}$Rn ($T_{1/2}$ = 3.8\,d) to diffuse into the sensitive volume near TU1. This effect is mitigated by the anti-radon box and its flushing. Nevertheless, $\gamma$ rays from the radon daughters $^{214}$Pb and $^{214}$Bi appear in the TU1 spectrum at 352 and 609\,keV, respectively. An additional weak feature at 295\,keV is not $2\sigma$ significant.

After a typical sample change, the integral counting rate is saturated at its final value after 10000\,s, which is significantly shorter than the physical half-life of  $^{222}$Rn, showing the desired effects of the active flushing of the anti-radon box. 

\subsubsection{Naturally occurring radionuclides}

The remaining rates in the $\gamma$-lines at 1460.8\,keV (from $^{40}$K) and 2614.5\,keV (from $^{208}$Tl) are 0.49(6)\,kg$^{-1}$d$^{-1}$ and 0.22(4)\,kg$^{-1}$d$^{-1}$ respectively (table \ref{Table:Gammalines}). Using the observed rates of these two $\gamma$ rays without shielding and applying a calculated attenuation based on the weakest spot of the passive shielding, counting rates of 0.3\,kg$^{-1}$d$^{-1}$ are found for these two lines. 

For $^{208}$Tl, based on these values there is no reason to assume a significant internal contamination of TU1 within the shielding materials or the detector.  

Also for $^{40}$K, the effect of the attenuation is comparable to the expectation. However, the attenuation is still smaller than conservatively assumed. The remaining rate of $^{40}$K within TU1 may therefore also include contaminations within the shielding and the detector. In fact, $^{40}$K has previously been found to be a remaining contaminant in high-purity copper samples \cite{XENON2017}. 


\section{Experimental study on $^7$Be}
\label{sec:7Be}

As an illustration for the capabilities of the TU1 detector for activation studies for nuclear astrophysics, a study of weak activated $^7$Be samples produced during the investigation of the $^3$He($\alpha,\gamma$)$^7$Be reaction at the Felsenkeller 5\,MV Pelletron accelerator is discussed here.

\subsection{Efficiency calibration}
\label{subsec:7BeEfficiency}

$^7$Be has a half-life of $T_{1/2}=53.22(6)$\,d and emits a $\gamma$-ray of $E=478\,$keV with an emission probability of $\eta=10.44(4)$\,\%. 

In order to determine the absolute full-energy peak efficiency at $E=478\,$keV, three calibration samples of $^7$Be have been produced at the cyclotron accelerator of the Institute for Nuclear Research in Debrecen (ATOMKI), Hungary. To this end, LiF has been evaporated on tantalum disks (0.22\,mm thickness and 27\,mm diameter) and subsequently bombarded with 5.5\,MeV protons, resulting in $^7$Be activities of 0.3, 22, and 48\,kBq, respectively. The geometry of the tantalum disks and proton beam spot has been chosen to conform to the geometry of the much less radioactive $^7$Be samples from the $^3$He($\alpha,\gamma$)$^7$Be irradiation. 

\begin{figure}[tb]
\centering
\includegraphics[width=\columnwidth,trim=0mm 0mm 15mm 15mm,clip]{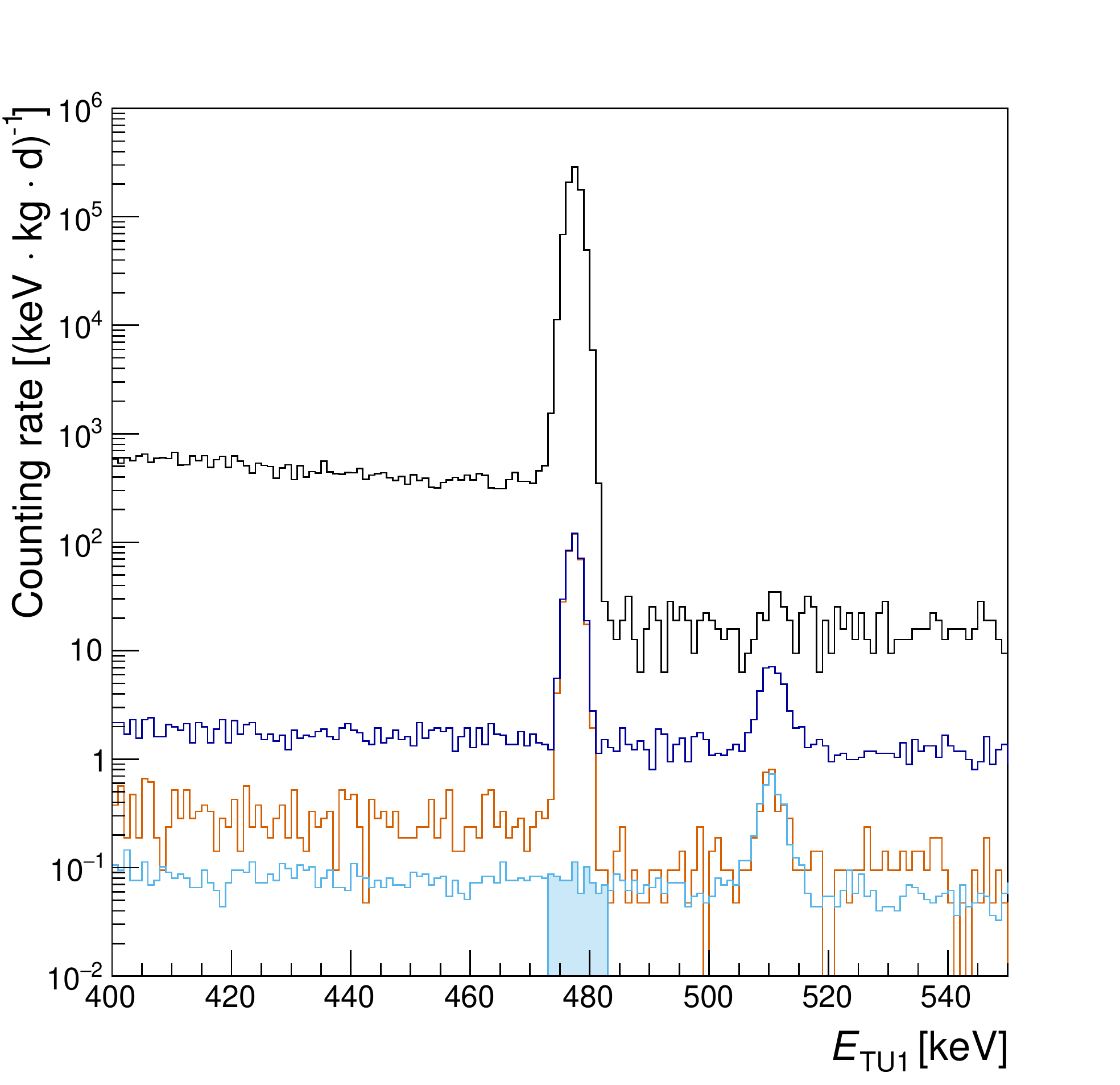}
\caption{Pulse height spectrum in the region of the 478\,keV line of $^7$Be. Black spectrum: $^7$Be calibration source with 2\,kBq. Blue: Activated $^7$Be sample (0.7\,Bq), without active veto. Red: Activated $^7$Be sample, with active veto. Light blue: background without sample, with active veto. Shaded light blue area: 10\,keV interval for the determination of the detection limit $L_\textrm{D}$. See text for details.}
\label{fig:7Be}
\end{figure}

The activity of the three ATOMKI $^7$Be samples has then determined in far geometry, using arrays of several other well-calibrated radionuclide standards, independently at ATOMKI and at HZDR. 
Using these three calibration sources, the absolute full-energy peak detection efficiency for the TU1 detector, when using a point-like sample in close geometry, has subsequently been determined to be $\varepsilon=14.45(19)\,\%$. 

\subsection{Activation analysis}

For the ongoing campaign on the $^3$He($\alpha,\gamma$)$^7$Be reaction at Felsenkeller, a tantalum disk (hereafter called sample ST8) of 0.22\,mm thickness and 27\,mm diameter has been implanted with 10$^{18}$ cm$^{-2}$ $^3$He at the HZDR Ion Beam Center. ST8 was then irradiated with a 5-10\,$\upmu$A beam of $^4$He$^+$ ($E_\alpha = 0.8-1.0\,$MeV) at Felsenkeller to investigate the $^3$He($\alpha,\gamma$)$^7$Be reaction, applying a charge of $Q=2.3\,$C.

The resulting pulse height spectrum near 478\,keV for the activated sample ST8 is shown in Figure \ref{fig:7Be}: blue curve without active veto, red curve with active veto. From the comparison of these two spectra, the cut survival probability of $P_{\rm cut \, survival}$ = 0.995 is found (Section \ref{subsec:EnergyCut}). The data are compared with the spectrum of the calibration sample (black curve) and the background without sample (light blue curve). 

\subsection{Sensitivity for $^7{\rm Be}$ detection}

\begin{table}[tb]
\caption{%
\label{Table:Sensitivity}%
Detection limit $L_\textrm{D}$ and activity $L_{\rm signal=noise}$ for signal equal to noise for 50\,d counting time and a 10\,keV wide region of interest for $^7$Be (478\,keV) and $^{181}$Hf (482\,keV). See text for details.} 
\begin{tabular}{lcc}
\hline 
Method & $^7$Be & $^{181}$Hf\\ 
& [mBq] & [mBq]\\
\hline
Detection limit $L_{\rm D}$ at 90\% CL & 0.81 & 0.11 \\
$L_{\rm signal=noise}$ & 2.20 & 0.31 \\
\hline
\end{tabular}
\end{table}

The nuclide $^7$Be is not only relevant for nuclear astrophysics \cite{Bemmerer2006,Orebi2021}. Due to its cosmogenic origin, it may also be used to monitor atmospherical and geological processes \cite{Terzi2020,Zhang2021,Tiessen18-JRNC}.

The present background spectrum (light blue curve in Figure \ref{fig:7Be}) and detection efficiency (section \ref{subsec:7BeEfficiency}), as well as  literature data on decay branching ratio and half-life, are now used in order to derive the sensitivity of TU1 for detecting $^7$Be.

The detection limit $L_{\rm D}$ in Bq, using a coverage factor of $k_\alpha$ = 1.282 for 90\% confidence level, a branching ratio $\beta$, an absolute full-enery peak detection efficiency $\varepsilon$, a background rate $\dot{N}_{\rm BG}$, and a counting time $t$ is given by \cite{Gilmore2008}
\begin{equation}
    L_{\rm D} = \frac{k_\alpha^2 + 2 k_\alpha\sqrt{2\dot{N}_{\rm BG}t}}{ \varepsilon \times \beta \times C_{\rm decay} \times t}.
\end{equation}
This value has been determined for a 10\,keV wide region of interest and two typical nuclides: In addition to $^7$Be, also for $^{181}$Hf, which has a similar half-life (42.39\,d) but a higher branching ratio, $\beta$ = 80.5\% for its 482\,keV $\gamma$ ray. Due to the similar half-lives, the decay corrections for 50\,d counting time are also similar, $C_{\rm decay}$ = 0.73 ($^7$Be) and 0.68 ($^{181}$Hf). Detection efficiency and background differ only negligibly for the two relevant energies, 478 and 482\,keV.

\begin{figure*}[bt]
\centering
\includegraphics[width=\textwidth,trim=7.7mm 0mm 15mm 5mm,clip]{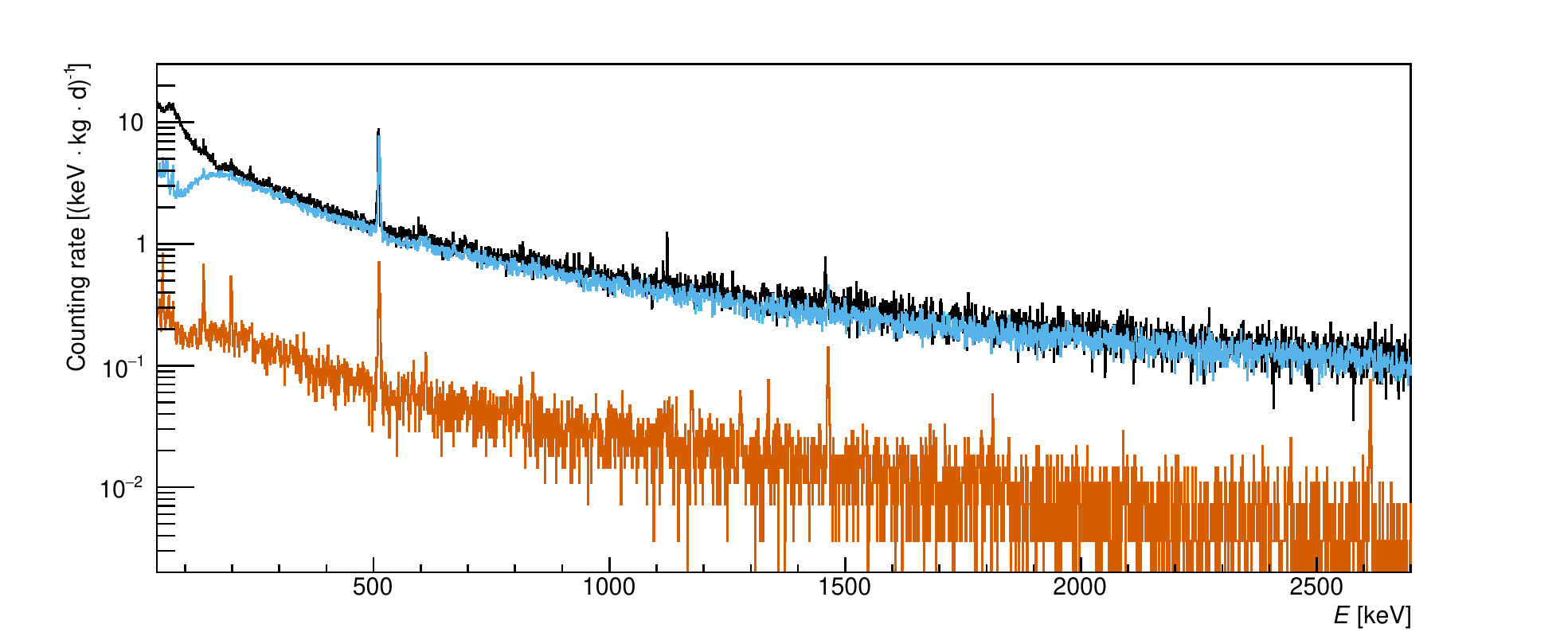}
\caption{Pulse height spectrum of the TU1 detector without active veto (blue spectrum) and with active veto (red spectrum), normalized to bin width and detector size and shown for the typical background region of interest within [40\,keV;2700\,keV]. For comparison, a previous spectrum from the D6 detector of the VKTA lab in Felsenkeller tunnel IV \cite{Koehler2009} is shown in black.
}
\label{fig:GammaSpectrumVKTA}
\end{figure*}

For comparison, for both cases also another value has been computed, namely the activity $L_{\rm signal=noise}$ for which the net count rate (signal) is equal to the background (noise). The data show that the TU1 detector can easily reach the $\upmu$Bq range, if a suitably long counting time is adopted (Table \ref{Table:Sensitivity}). 

The same background data was then used to calculate the maximum half-life $T_{1/2}^{\rm max}$ that can be determined in a sample with $N_{\rm A}$ (1\,mol) decaying nuclei, neglecting the decay and self-absorption corrections. Again 50\,d counting time, a 10\,keV wide region of interest near 478\,keV, and a detection efficiency of 14.45\% are used, and now a branching of $\beta$ = 100\%.
\begin{eqnarray}
        T_{1/2}^{\rm max} & = & \frac{\ln 2}{\lambda^{\rm max}} = \frac{\ln 2 \times N_{\rm A}}{L_{\rm D}} \nonumber \\
        & = & \frac{\ln 2 \times N_{\rm A} \times \varepsilon \times \beta \times C_{\rm decay} \times t}{k_\alpha^2 + 2 k_\alpha\sqrt{2\dot{N}_{\rm BG}t}} \\
        & = & 2.1 \times 10 ^{20} \textrm{y} \nonumber 
\end{eqnarray}
This value is in the range of double-beta decay half-lives with the emission of neutrinos that have recently been determined \cite{Zuber2021}.

\section{Discussion}
\label{sec:Discussion}

In this section, the new data are interpreted and compared with literature data. 

\subsection{Interpretation of the present data}

The passive shielding suppresses the integrated counting rate between 40\,keV and 2700\,keV by a factor of 4300 (Figures \ref{fig:TU1_spectrum} and \ref{fig:GammaSpectrumVKTA}), showing its high effectivity. In the passively shielded spectrum (blue curve in Figure \ref{fig:GammaSpectrumVKTA}), only two peaks from the natural background remain, namely $^{40}$K and $^{208}$Tl. The rest of the passively shielded spectrum is dominated by muon-induced events, both in the continuum and in the 511\,keV annihilation peak. 

Furthermore, the passively shielded spectrum is very similar to a previously published spectrum \cite{Koehler2009} that has been obtained in the VKTA laboratory in the nearby Felsenkeller tunnel IV, which has the same rock overburden and muon flux as tunnel VIII studied here \cite{Ludwig2019}. There are only two significant differences between these spectra (black and blue curves in Figure \ref{fig:GammaSpectrumVKTA}). 

First, a slightly lower $^{40}$K rate in the present spectrum. This is remarkable, given that there is a significant amount of $^{40}$K in the laboratory walls surrounding TU1 (section \ref{subsec:PassiveShielding}). In the case of tunnel IV, the $^{40}$K $\gamma$ rays are already attenuated outside the detector shield itself by a measurement chamber made of ancient steel (MK2) that hosts several detectors \cite{Koehler2009}. The comparison seems to indicate that the present passive shield, 15\,cm of radiopure lead and 10\,cm of copper, improves upon the previous one (10\,cm normal lead, 5\,cm radiopure lead, 5\,cm radiopure copper) \cite{Koehler2009}. Importantly, it shows that the omission of a measurement chamber like MK2 is more than compensated by the present, additional 5\,cm of copper and, possibly, by the usage of radiopure lead for the outer shielding. The second difference between the black and blue spectra is a $^{65}$Zn contamination ($E_\gamma$ = 1115\,keV, $T_{1/2}$ = 244\,d) in the VKTA spectrum, which has been measured several years ago. While this contamination used to be significant, in the meantime it decayed. 

The present active shielding further reduces the integrated counting rate by a factor of 17, bringing the total background suppression to a factor of 73000, almost five orders of magnitude. At the same time, random coincidences reduce the peak detection efficiency by just 0.47\%. This small reduction is acceptable given the fact that the setup is designed for low counting rate experiments that will be limited by the statistical, not the systematical uncertainty.

\begin{table*}[ht]
\caption{%
\label{Table:Comparison}%
Background count rates for selected detectors, normalized to detector mass. The energy interval is [40\,keV;2700\,keV], except for the Gator detector at LNGS where the energy interval starts only at 60 keV.}
\begin{tabular}{llrD{;}{\pm}{5}D{;}{\pm}{4}l}
\hline 
Detector/ & Location & \multicolumn{1}{l}{Depth} & \multicolumn{1}{l}{Count rate} & \multicolumn{1}{l}{Count rate} & Reference\\ 
Laboratory & & \multicolumn{1}{l}{[m.w.e]} & \multicolumn{1}{l}{without veto} & \multicolumn{1}{l}{with veto} \\
& & & \multicolumn{1}{l}{[kg$^{-1}$ d$^{-1}$]} & \multicolumn{1}{l}{[kg$^{-1}$ d$^{-1}$]} & \\
\hline
D4 & Seibersdorf Laboratory, Austria & $\approx$1 & 85100;400 & 8110;40 & \cite{Schroettner2022, Schwaiger2002}\\
DLB & TU Dortmund, Germany & 10 &34400;60 &2900;6 & \cite{Nitsch2017} \\
GIOVE & MPIK Heidelberg, Germany & 15 & 31027;48 & 348;3 & \cite{Heusser2015, Hakenmueller2015}\\
CAVE & IAEA, Monaco & 54\footnotemark &  & 840;50 & \cite{Laubenstein2004}\\
D6 & Felsenkeller (VKTA), Germany & 110 & 2938;5 & \multicolumn{1}{c}{} & \cite{Koehler2009}\\
TU1 & Felsenkeller (TU Dresden and HZDR), Germany & 140 & 1982;3 & 116;1 & present work\\
Ge-14 & HADES, Belgium & 500 & 208;4 & 178;8 & \cite{Hult2018, Hult2022}\\
GeMSE & La Vue des Alpes Laboratory, Switzerland & 620 & \multicolumn{1}{c}{} & 91;1 & \cite{Garcia2022, PerComGarcia2022}\\
GeOroel & LSC, Canfranc, Spain & 2450 & \multicolumn{1}{D{;}{\,}{6}}{142;} & \multicolumn{1}{c}{} & \cite{Perez2021} \\
GeCRIS & LNGS, Gran Sasso, Italy & 3800 & 111;1  &\multicolumn{1}{c}{} & \cite{Laubenstein2022} \\ 
GeMPI & LNGS, Gran Sasso, Italy & 3800 & 59;1 & \multicolumn{1}{c}{} & \cite{Laubenstein2022}\\ 
Gator & LNGS, Gran Sasso, Italy -- [60 keV;2700 keV] & 3800 & 89.0;0.7  &\multicolumn{1}{c}{} & \cite{Araujo2022} \\
OBELIX & LSM, Modane, France & 4800 & 68;1 & \multicolumn{1}{c}{} & \cite{Brudanin2017} \\
\hline
\end{tabular}
\end{table*}

The active shield reveals a number of $\gamma$ rays that had been covered by the muon-induced continuum in the spectrum without active shield. It is possible that a future slight increase of the lead shield thickness, by up to an additional 5\,cm, might further attenuate the remaining radionuclide lines from $^{40}$K and $^{208}$Tl, at the expense of a somewhat higher ($\mu,n$) rate \cite{Grieger2020} that would increase the neutron-induced lines at 198\,keV and below (sections \ref{subsec:NeutronInducedPrompt} and \ref{subsec:NeutronInducedActivation}).

The actively shielded spectrum still shows a significant continuum, without any clearly discernible Compton edges (Figure \ref{fig:GammaSpectrumVKTA}, red spectrum). This may indicate that correcting the small imperfections in the active muon veto, for example the holes for the lifting mechanism of the lid, may lead to a further background reduction. In another shallow-underground laboratory, the muon veto reduced the 40-2700 keV counting rate by a factor of 89 \cite{Heusser2015}, indicating that correcting the small imperfections in the muon veto may yield another factor of five reduction in counting rate.

\subsection{Comparison with other setups reported in the literature}
\label{subsec:Comparison}

\begin{figure}[t]
\centering
\includegraphics[width=\columnwidth,trim=0mm 0mm 0mm 0mm,clip]{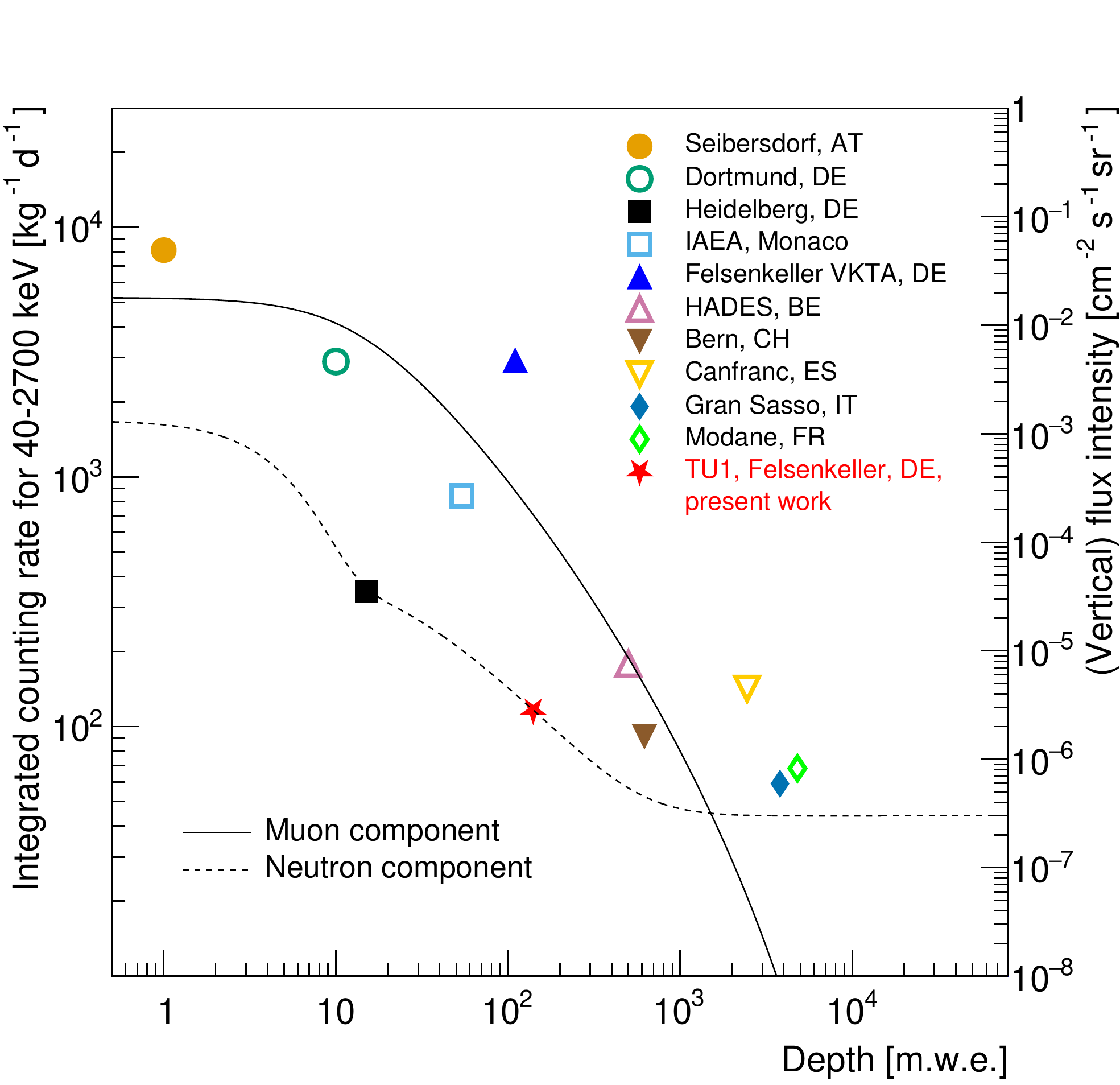}
\caption{Integrated counting rates in the [40\,keV;2700\,keV] region for different low-background detectors as a function of the rock overburden. Data from Table \ref{Table:Comparison} and with GeMPI plotted in case of the three LNGS detectors. The vertical flux intensity of muons and the flux intensity neutrons are estimated and added on the second y-axis \cite{Barbouti1983,Grieger2021}. See text for details. 
}
\label{fig:RateComparison}
\end{figure}

In order to further extend the comparison and discussion, data from a selection of low-background HPGe detectors in underground laboratories is listed in table \ref{Table:Comparison}. Their integrated counting rates within the energy interval of [40\,keV,2700\,keV] (for the Gran Sasso detectors, a slightly different range of [100\,keV,2700\,keV] is used) are plotted in figure \ref{fig:RateComparison} as a function of their corresponding rock overburden.

The setups used for comparison span a broad range of rock overburden and generally reflect the state of the art, with expert adjustments being made for the relevant depths. 

Two general trends are apparent (Figure \ref{fig:RateComparison}): First, the counting rate generally decreases with increasing rock overburden. Second, the scatter of the data points from different laboratories decreases with increasing rock overburden. Both of these effects are due to cosmic-ray induced muons. Once the radionuclide-induced background has been strongly attenuated, the muon flux remains the main variable affecting the observed integral background rate. While at high rock overburden, the muon flux is attenuated so much that it plays no major role any more, at medium-low rock overburden, it is important to optimize the treatment of muon-induced effects, including an active veto. The differences in the particular treatment of the muon background in various laboratories are reflected in the spread of counting rates at similar depths (Figure \ref{fig:RateComparison}).

It is apparent that the background rate in the present setup is in the same league with much deeper underground detectors, and that it is lower than the background of some setups with even greater rock overburden.

\section{Summary and outlook}
\label{sec:Conclusion}

The present work describes a recently installed new HPGe detector called TU1 in the shallow-underground laboratory Felsenkeller (Dresden, Germany). Using a sophisticated passive and active shield, the integrated background counting rate in the detector is reduced to 116\,kg$^{-1}$d$^{-1}$ in the [40\,keV;2700\,keV] interval, an unprecedented low value for a shallow-underground laboratory. This detector is now the most sensitive radioactivity-measurement setup in Germany. 

A detailed study of the remaining background is presented, and for the examples of point-like sources of $^7$Be and $^{181}$Hf, the detection limits are derived for typical counting times. The data are compared to relevant similar laboratories. 

Comprehensive simulation studies are currently ongoing and with more statistics in the coming years, it will be possible to explore the correlations between incoming muons and their signature in the shielded germanium detector in greater detail. At the same time, it is planned to further improve muon veto efficiency by closing remaining small gaps in the active scintillator shield.  


\subsection*{Acknowledgments}
	
The authors are indebted to Tamás Szücs (ATOMKI Debrecen) for providing the $^7$Be calibration samples, to Detlev Degering (VKTA) for valuable discussions, and to Toralf Döring, Maik Görler, Andreas Hartmann, Bernd Rimarzig (HZDR), and Martin Siegel (TU Dresden) for technical support. ---
Financial support by Deutsche Forschungsgemeinschaft DFG (INST 269/631-1 FUGG, TU Dresden Institutional Strategy ”support the best”, ZU123/21-1, and BE4100/4-1), by the Konrad-Adenauer-Stiftung, and by the European Union (ChETEC-INFRA, project no. 101008324) is gratefully acknowledged.

\footnotetext[3]{Due to the geometric complexity of the rock overburden, the effective depth of this laboratory was calculated based on the integrated muon intensity. \cite{Ludwig2021, Barbouti1983}}



\begin{thebibliography}{10}
\expandafter\ifx\csname url\endcsname\relax
  \def\url#1{\texttt{#1}}\fi
\expandafter\ifx\csname urlprefix\endcsname\relax\def\urlprefix{URL }\fi
\expandafter\ifx\csname href\endcsname\relax
  \def\href#1#2{#2} \def\path#1{#1}\fi

\bibitem{Povinec07-Book}
P.~Povinec (Ed.), Analysis of Environmental Radionuclides (Radioactivity in the
  Environment), Vol.~11, Elsevier Science, 2007.

\bibitem{SuperCDMS2017}
R.~Agnese, A.~Anderson, T.~Aramaki, I.~Arnquist, W.~Baker, D.~Barker, R.~B.
  Thakur, D.~Bauer, A.~Borgland, M.~Bowles, et~al., Projected sensitivity of
  the {S}uper{CDMS} {SNOLAB} experiment, Physical Review D 95~(8) (2017)
  082002.

\bibitem{Edelweiss2017}
E.~Armengaud, Q.~Arnaud, C.~Augier, A.~Beno{\^\i}t, L.~Berg{\'e}, T.~Bergmann,
  J.~Billard, T.~De~Boissi{\`e}re, G.~Bres, A.~Broniatowski, et~al.,
  Performance of the {EDELWEISS}-{III} experiment for direct dark matter
  searches, Journal of Instrumentation 12~(08) (2017) P08010.

\bibitem{Zuber2021}
K.~Zuber, {Neutrino Physics, Series in High Energy Physics, Cosmology and
  Gravitation}, CRC Press, 2021.

\bibitem{Brudanin2017}
V.~Brudanin, V.~Egorov, R.~Hodak, A.~Klimenko, P.~Loaiza, F.~Mamedov,
  F.~Piquemal, E.~Rukhadze, N.~Rukhadze, I.~{\v{S}}tekl, et~al., Development of
  the ultra-low background {HPG}e spectrometer {OBELIX} at {Modane} underground
  laboratory, Journal of Instrumentation 12~(02) (2017) P02004.

\bibitem{Alvis2019}
S.~Alvis, I.~Arnquist, F.~Avignone~III, A.~Barabash, C.~Barton, V.~Basu,
  F.~Bertrand, B.~Bos, M.~Busch, M.~Buuck, et~al., Search for neutrinoless
  double-$\beta$ decay in $^{76}${G}e with 26 kg yr of exposure from the
  {Majorana} demonstrator, Physical Review C 100~(2) (2019) 025501.

\bibitem{2020PhRvL.125y2502A}
M.~{Agostini}, G.~R. {Araujo}, A.~M. {Bakalyarov}, M.~{Balata}, I.~{Barabanov},
  L.~{Baudis}, C.~{Bauer}, E.~{Bellotti}, S.~{Belogurov}, A.~{Bettini},
  L.~{Bezrukov}, V.~{Biancacci}, D.~{Borowicz}, E.~{Bossio}, V.~{Bothe},
  V.~{Brudanin}, R.~{Brugnera}, A.~{Caldwell}, C.~{Cattadori},
  A.~{Chernogorov}, T.~{Comellato}, V.~{D'Andrea}, E.~V. {Demidova}, N.~{di
  Marco}, E.~{Doroshkevich}, F.~{Fischer}, M.~{Fomina}, A.~{Gangapshev},
  A.~{Garfagnini}, C.~{Gooch}, P.~{Grabmayr}, V.~{Gurentsov}, K.~{Gusev},
  J.~{Hakenm{\"u}ller}, S.~{Hemmer}, R.~{Hiller}, W.~{Hofmann}, J.~{Huang},
  M.~{Hult}, L.~V. {Inzhechik}, J.~{Janicsk{\'o} Cs{\'a}thy}, J.~{Jochum},
  M.~{Junker}, V.~{Kazalov}, Y.~{Kerma{\"\i}dic}, H.~{Khushbakht}, T.~{Kihm},
  I.~V. {Kirpichnikov}, A.~{Klimenko}, R.~{Knei{\ss}l}, K.~T. {Kn{\"o}pfle},
  O.~{Kochetov}, V.~N. {Kornoukhov}, P.~{Krause}, V.~V. {Kuzminov},
  M.~{Laubenstein}, A.~{Lazzaro}, M.~{Lindner}, I.~{Lippi}, A.~{Lubashevskiy},
  B.~{Lubsandorzhiev}, G.~{Lutter}, C.~{Macolino}, B.~{Majorovits},
  W.~{Maneschg}, L.~{Manzanillas}, M.~{Miloradovic}, R.~{Mingazheva},
  M.~{Misiaszek}, P.~{Moseev}, Y.~{M{\"u}ller}, I.~{Nemchenok}, K.~{Panas},
  L.~{Pandola}, K.~{Pelczar}, L.~{Pertoldi}, P.~{Piseri}, A.~{Pullia},
  C.~{Ransom}, L.~{Rauscher}, S.~{Riboldi}, N.~{Rumyantseva}, C.~{Sada},
  F.~{Salamida}, S.~{Sch{\"o}nert}, J.~{Schreiner}, M.~{Sch{\"u}tt}, A.~K.
  {Sch{\"u}tz}, O.~{Schulz}, M.~{Schwarz}, B.~{Schwingenheuer},
  O.~{Selivanenko}, E.~{Shevchik}, M.~{Shirchenko}, L.~{Shtembari},
  H.~{Simgen}, A.~{Smolnikov}, D.~{Stukov}, A.~A. {Vasenko}, A.~{Veresnikova},
  C.~{Vignoli}, K.~{von Sturm}, T.~{Wester}, C.~{Wiesinger}, M.~{Wojcik},
  E.~{Yanovich}, B.~{Zatschler}, I.~{Zhitnikov}, S.~V. {Zhukov},
  D.~{Zinatulina}, A.~{Zschocke}, A.~J. {Zsigmond}, K.~{Zuber}, G.~{Zuzel},
  {Gerda Collaboration}, {Final Results of GERDA on the Search for Neutrinoless
  Double-{\ensuremath{\beta}} Decay}, Phys. Rev. Lett. 125~(25) (2020) 252502.
\newblock \href {https://doi.org/10.1103/PhysRevLett.125.252502}
  {\path{doi:10.1103/PhysRevLett.125.252502}}.

\bibitem{Laubenstein2004}
M.~Laubenstein, M.~Hult, J.~Gasparro, D.~Arnold, S.~Neumaier, G.~Heusser,
  M.~K{\"o}hler, P.~Povinec, J.-L. Reyss, M.~Schwaiger, P.~Theodórssoni,
  Underground measurements of radioactivity, Applied Radiation and Isotopes
  61~(2-3) (2004) 167--172.

\bibitem{Arpesella96-Apradiso}
C.~Arpesella, A low background counting facility at {L}aboratori {N}azionali
  del {G}ran {S}asso., Appl. Radiat. Isot. 47 (1996) 991--996.

\bibitem{Ludwig2019}
F.~Ludwig, L.~Wagner, T.~Al-Abdullah, G.~Barnaf{\"o}ldi, D.~Bemmerer,
  D.~Degering, K.~Schmidt, G.~Sur{\'a}nyi, T.~Sz{\"u}cs, K.~Zuber, The muon
  intensity in the {F}elsenkeller shallow underground laboratory, Astroparticle
  Physics 112 (2019) 24--34.

\bibitem{Szuecs2019}
T.~Sz{\"u}cs, D.~Bemmerer, D.~Degering, A.~Domula, M.~Grieger, F.~Ludwig,
  K.~Schmidt, J.~Steckling, S.~Turkat, K.~Zuber, Background in $\gamma$-ray
  detectors and carbon beam tests in the {F}elsenkeller shallow-underground
  accelerator laboratory, The European Physical Journal A 55~(10) (2019) 1--12.

\bibitem{Grieger2020}
M.~Grieger, T.~Hensel, J.~Agramunt, D.~Bemmerer, D.~Degering, I.~Dillmann,
  L.~Fraile, D.~Jordan, U.~K{\"o}ster, M.~Marta, et~al., Neutron flux and
  spectrum in the {D}resden {F}elsenkeller underground facility studied by
  moderated {H}e3 counters, Physical Review D 101~(12) (2020) 123027.

\bibitem{Wulandari04-APP}
H.~{Wulandari}, J.~{Jochum}, W.~{Rau}, F.~{von Feilitzsch}, {Neutron flux at
  the Gran Sasso underground laboratory revisited}, Astropart.~Phys. 22 (2004)
  313--322.

\bibitem{Koehler2009}
M.~K{\"o}hler, D.~Degering, M.~Laubenstein, P.~Quirin, M.-O. Lampert, M.~Hult,
  D.~Arnold, S.~Neumaier, J.-L. Reyss, A new low-level $\gamma$-ray
  spectrometry system for environmental radioactivity at the underground
  laboratory {F}elsenkeller, Applied Radiation and Isotopes 67~(5) (2009)
  736--740.

\bibitem{Bemmerer2006}
D.~Bemmerer, F.~Confortola, H.~Costantini, A.~Formicola, G.~Gy{\"u}rky,
  R.~Bonetti, C.~Broggini, P.~Corvisiero, Z.~Elekes, Z.~F{\"u}l{\"o}p, et~al.,
  Activation measurement of the $^3${H}e($\alpha,\gamma$)$^7${B}e cross section
  at low energy, Physical {R}eview {L}etters 97~(12) (2006) 122502.

\bibitem{2014PhRvC..89a5803D}
A.~{Di Leva}, D.~A. {Scott}, A.~{Caciolli}, A.~{Formicola}, F.~{Strieder},
  M.~{Aliotta}, M.~{Anders}, D.~{Bemmerer}, C.~{Broggini}, P.~{Corvisiero},
  Z.~{Elekes}, Z.~{F{\"u}l{\"o}p}, G.~{Gervino}, A.~{Guglielmetti},
  C.~{Gustavino}, G.~{Gy{\"u}rky}, G.~{Imbriani}, J.~{Jos{\'e}}, M.~{Junker},
  M.~{Laubenstein}, R.~{Menegazzo}, E.~{Napolitani}, P.~{Prati}, V.~{Rigato},
  V.~{Roca}, E.~{Somorjai}, C.~{Salvo}, O.~{Straniero}, T.~{Sz{\"u}cs},
  F.~{Terrasi}, D.~{Trezzi}, {LUNA Collaboration}, {Underground study of the
  $^{17}$O(p,{\ensuremath{\gamma}})$^{18}$F reaction relevant for explosive
  hydrogen burning}, Phys. Rev. C 89~(1) (2014) 015803.
\newblock \href {https://doi.org/10.1103/PhysRevC.89.015803}
  {\path{doi:10.1103/PhysRevC.89.015803}}.

\bibitem{2013PhRvC..88b5803S}
K.~{Schmidt}, S.~{Akhmadaliev}, M.~{Anders}, D.~{Bemmerer}, K.~{Boretzky},
  A.~{Caciolli}, D.~{Degering}, M.~{Dietz}, R.~{Dressler}, Z.~{Elekes},
  Z.~{F{\"u}l{\"o}p}, G.~{Gy{\"u}rky}, R.~{Hannaske}, A.~R. {Junghans},
  M.~{Marta}, M.-L. {Menzel}, F.~{Munnik}, D.~{Schumann}, R.~{Schwengner},
  T.~{Sz{\"u}cs}, A.~{Wagner}, D.~{Yakorev}, K.~{Zuber}, {Resonance triplet at
  E$_{{\ensuremath{\alpha}}}$=4.5 MeV in the
  $^{40}$Ca({\ensuremath{\alpha}},{\ensuremath{\gamma}})$^{44}$Ti reaction},
  Phys. Rev. C 88~(2) (2013) 025803.
\newblock \href {http://arxiv.org/abs/1307.6516} {\path{arXiv:1307.6516}},
  \href {https://doi.org/10.1103/PhysRevC.88.025803}
  {\path{doi:10.1103/PhysRevC.88.025803}}.

\bibitem{2018PrPNP..98...55B}
C.~{Broggini}, D.~{Bemmerer}, A.~{Caciolli}, D.~{Trezzi}, {LUNA: Status and
  prospects}, Progress in Particle and Nuclear Physics 98 (2018) 55--84.
\newblock \href {http://arxiv.org/abs/1707.07952} {\path{arXiv:1707.07952}},
  \href {https://doi.org/10.1016/j.ppnp.2017.09.002}
  {\path{doi:10.1016/j.ppnp.2017.09.002}}.

\bibitem{Orebi2021}
G.~D. Orebi~Gann, K.~Zuber, D.~Bemmerer, A.~Serenelli, The future of solar
  neutrinos, Annual Review of Nuclear and Particle Science 71 (2021) 491--528.

\bibitem{Anders2014}
M.~Anders, D.~Trezzi, R.~Menegazzo, M.~Aliotta, A.~Bellini, D.~Bemmerer,
  C.~Broggini, A.~Caciolli, P.~Corvisiero, H.~Costantini, et~al., First direct
  measurement of the $^2${H}($\alpha,\gamma$)$^6${L}i cross section at big bang
  energies and the primordial lithium problem, Physical {R}eview {L}etters
  113~(4) (2014) 042501.

\bibitem{Mossa2020}
V.~Mossa, K.~St{\"o}ckel, F.~Cavanna, F.~Ferraro, M.~Aliotta, F.~Barile,
  D.~Bemmerer, A.~Best, A.~Boeltzig, C.~Broggini, et~al., The baryon density of
  the universe from an improved rate of deuterium burning, Nature 587~(7833)
  (2020) 210--213.

\bibitem{Turkat2021}
S.~Turkat, S.~Hammer, E.~Masha, S.~Akhmadaliev, D.~Bemmerer, M.~Grieger,
  T.~Hensel, J.~Julin, M.~Koppitz, F.~Ludwig, et~al., Measurement of the
  $^2${H}(p,$\gamma$)$^3${H}e {S} factor at 265-1094 ke{V}, Physical Review C
  103~(4) (2021) 045805.

\bibitem{Paelchen2008}
W.~P{\"a}lchen, H.~Walter, Geologie von {S}achsen, E. Schweizerbart'sche
  Verlagsbuchhandlung, 2008.

\bibitem{Grieger2021}
M.~Grieger,
  \href{https://nbn-resolving.org/urn:nbn:de:bsz:14-qucosa2-776845}{Neutronenfluss
  in {U}ntertagelaboren}, Ph.D. thesis, TU Dresden (2021).
\newline\urlprefix\url{https://nbn-resolving.org/urn:nbn:de:bsz:14-qucosa2-776845}

\bibitem{Stekl2021}
I.~{\v{S}}tekl, J.~Hůlka, F.~Mamedov, P.~Fojt{\'\i}k,
  E.~{\v{C}}erm{\'a}kov{\'a}, K.~J{\'\i}lek, M.~Havelka, R.~Hod{\'a}k,
  M.~H{\`y}{\v{z}}a, Low radon cleanroom for underground laboratories,
  Frontiers in Public Health 8 (2021) 1086.

\bibitem{Reinhardt16-NIMA}
T.~P. {Reinhardt}, S.~{Gohl}, S.~{Reinicke}, D.~{Bemmerer}, T.~E. {Cowan},
  K.~{Heidel}, M.~{R{\"o}der}, D.~{Stach}, A.~{Wagner}, D.~{Weinberger},
  K.~{Zuber}, {for the R3B collaboration}, {Silicon photomultiplier readout of
  a monolithic 270$\times$5$\times$5 cm$^3$ plastic scintillator bar for time
  of flight applications}, Nucl.~Inst.~Meth.~A 816 (2016) 16--24.

\bibitem{Groom2001}
D.~E. Groom, N.~V. Mokhov, S.~I. Striganov, Muon stopping power and range
  tables 10 {M}e{V}--100 {T}e{V}, Atomic Data and Nuclear Data Tables 78~(2)
  (2001) 183--356.

\bibitem{CAEN}
Caen {S.p.A.}, \url{https://www.caen.it} ({Accessed: 2021-12-22}).

\bibitem{Bunting74-NIM}
R.~{Bunting}, J.~J. Kraushaar, {Short-lived radioactivity induced in Ge(Li)
  gamma-ray detectors by neutrons}, Nucl.~Inst.~Meth. 118 (1974) 565--572.

\bibitem{Skoro1992}
G.~{\v{S}}koro, I.~Ani{\v{c}}in, A.~Kuko{\v{c}}, D.~Krmpoti{\'c},
  P.~Ad{\v{z}}i{\'c}, R.~Vukanovi{\'c}, M.~{\v{Z}}upan{\v{c}}i{\'c},
  Environmental neutrons as seen by a germanium gamma-ray spectrometer, Nuclear
  Instruments and Methods in Physics Research Section A: Accelerators,
  Spectrometers, Detectors and Associated Equipment 316~(2-3) (1992) 333--336.

\bibitem{Hensel19-Master}
T.~Hensel, {Messung des nat{\"u}rlichen Neutronenspektrums unter Tage bei
  niedrigem Untergrund}, Master's thesis, TU Dresden (2019).

\bibitem{Heusser93-NIMB}
G.~Heusser, {Cosmic-ray induced background in Ge-spectrometry.},
  Nucl.~Inst.~Meth.~B 83 (1993) 223--228.

\bibitem{Anders2013}
M.~Anders, D.~Bemmerer, Z.~Elekes, M.~Marta, D.~Trezzi, C.~Mazzocchi,
  A.~Bellini, H.~Costantini, P.~Corvisiero, A.~Lemut, et~al., Neutron-induced
  background by an $\alpha$-beam incident on a deuterium gas target and its
  implications for the study of the $^2${H}($\alpha$,$\gamma$)$^6${L}i reaction
  at {LUNA}, The European Physical Journal, A 49 (2013).

\bibitem{Abusaleem2011}
K.~Abusaleem, B.~Singh, Nuclear data sheets for {A}=71, Nuclear Data Sheets
  112~(1) (2011) 133--273.

\bibitem{XENON2017}
{{XENON} Collaboration and Elena Aprile and others}, Material radioassay and
  selection for the {XENON1T} dark matter experiment, European Physical Journal
  C 77~(12) (2017) 890.
\newblock \href {https://doi.org/10.1140/epjc/s10052-017-5329-0}
  {\path{doi:10.1140/epjc/s10052-017-5329-0}}.

\bibitem{Terzi2020}
L.~Terzi, G.~Wotawa, M.~Schoeppner, M.~Kalinowski, P.~R. Saey, P.~Steinmann,
  L.~Luan, P.~W. Staten, Radioisotopes demonstrate changes in global
  atmospheric circulation possibly caused by global warming, Scientific reports
  10~(1) (2020) 1--13.

\bibitem{Zhang2021}
F.~Zhang, J.~Wang, M.~Baskaran, Q.~Zhong, Y.~Wang, J.~Paatero, J.~Du, A
  comprehensive global dataset of atmospheric $^{7}${B}e and $^{210}${P}b
  measurements: air concentration and depositional flux, Earth System Science
  Data 7 (2021) 1--75.

\bibitem{Tiessen18-JRNC}
C.~Tiessen, D.~Bemmerer, G.~Rugel, R.~Querfeld, A.~Scharf, G.~Steinhauser,
  S.~Merchel, {Accelerator mass spectrometry (AMS) for beryllium-7 measurements
  in smallest rainwater samples}, Journal of Radioanalytical and Nuclear
  Chemistry 319 (2019) 975--973.

\bibitem{Gilmore2008}
G.~Gilmore, Practical gamma-ray spectroscopy, John Wiley \& Sons, 2008.

\bibitem{Schroettner2022}
{P}ersonal communication with {T}. {S}chröttner, {S}eibersdorf {L}abor {GmbH},
  2444 {S}eibersdorf, {A}ustria. (2022).

\bibitem{Schwaiger2002}
M.~Schwaiger, F.~Steger, T.~Schroettner, C.~Schmitzer, A ultra low level
  laboratory for nuclear test ban measurements, Applied radiation and isotopes
  56~(1-2) (2002) 375--378.

\bibitem{Nitsch2017}
C.~Nitsch, M.~Gerhardt, C.~G{\"o}{\ss}ling, K.~Kr{\"o}ninger, Improvements to
  the muon veto of the {D}ortmund {L}ow {B}ackground {F}acility, Applied
  Radiation and Isotopes 126 (2017) 201--203.

\bibitem{Heusser2015}
G.~Heusser, M.~Weber, J.~Hakenm{\"u}ller, M.~Laubenstein, M.~Lindner,
  W.~Maneschg, H.~Simgen, D.~Stolzenburg, H.~Strecker, {GIOVE}: a new detector
  setup for high sensitivity germanium spectroscopy at shallow depth, The
  European Physical Journal C 75~(11) (2015) 1--16.

\bibitem{Hakenmueller2015}
J.~Hakenmüller, Simulation of the cosmic ray induced background in the {GIOVE}
  detector, Master's thesis, University of Heidelberg (2015).

\bibitem{Hult2018}
M.~Hult, G.~Marissens, H.~Stroh, G.~Lutter, F.~Tzika, N.~Markovi{\'c},
  Characterisation of an ultra low-background point contact {HPG}e
  well-detector for an underground laboratory, Applied Radiation and Isotopes
  134 (2018) 446--449.

\bibitem{Hult2022}
{P}ersonal communication with {M}. {H}ult (2022).

\bibitem{Garcia2022}
D.~{Ram{\'\i}rez Garc{\'\i}a}, D.~{Baur}, J.~{Grigat}, B.~A. {Hofmann},
  S.~{Lindemann}, D.~{Masson}, M.~{Schumann}, M.~{von Sivers}, F.~{Toschi},
  {GeMSE: a low-background facility for gamma-spectrometry at moderate rock
  overburden}, Journal of Instrumentation 17~(4) (2022) P04005.
\newblock \href {http://arxiv.org/abs/2202.06540} {\path{arXiv:2202.06540}},
  \href {https://doi.org/10.1088/1748-0221/17/04/P04005}
  {\path{doi:10.1088/1748-0221/17/04/P04005}}.

\bibitem{PerComGarcia2022}
{P}ersonal communication with {D}. {R}. {G}arcía and {M}. {S}chumann (2022).

\bibitem{Perez2021}
J.~P{\'e}rez-P{\'e}rez, J.~C. Amare, I.~C. Bandac, A.~Bayo,
  S.~Borjabad-Sanchez, J.~M. Calvo-Mozota, L.~Cid-Barrio,
  R.~Hern{\'a}ndez-Antol{\'\i}n, B.~Hern{\'a}ndez-Molinero, P.~Novella, et~al.,
  Radon mitigation applications at the {L}aboratorio {S}ubterr{\'a}neo de
  {C}anfranc ({LSC}), Universe 8~(2) (2022) 112.

\bibitem{Laubenstein2022}
{P}ersonal communication with {M}. {L}aubenstein (2022).

\bibitem{Araujo2022}
G.~R. {Araujo}, L.~{Baudis}, Y.~{Biondi}, A.~{Bismark}, M.~{Galloway}, {The
  upgraded low-background germanium counting facility Gator for
  high-sensitivity {\ensuremath{\gamma}}-ray spectrometry}, Journal of
  Instrumentation 17~(8) (2022) P08010.
\newblock \href {http://arxiv.org/abs/2204.12478} {\path{arXiv:2204.12478}},
  \href {https://doi.org/10.1088/1748-0221/17/08/P08010}
  {\path{doi:10.1088/1748-0221/17/08/P08010}}.

\bibitem{Barbouti1983}
A.~Barbouti, B.~Rastin, A study of the absolute intensity of muons at sea level
  and under various thicknesses of absorber, Journal of Physics G: Nuclear
  Physics 9~(12) (1983) 1577.

\bibitem{Ludwig2021}
F.~Ludwig, Underground measurements and simulations on the muon intensity and
  $^{12}${C}-induced nuclear reactions at low energies, Ph.D. thesis, TU
  Dresden (2021).

\end{thebibliography}
\end{document}